\documentclass[notitlepage,aps,prd,twocolumn,reprint,superscriptaddress,showpacs,floatfix]{revtex4-1}

\usepackage{natbib}
\usepackage{graphicx}

\usepackage{amsmath, amssymb}
\usepackage{mathtools}
\usepackage{siunitx}
\usepackage[export]{adjustbox}

\usepackage{subfig}
\usepackage{xfrac}
\usepackage{multirow}
\usepackage{titlesec}

\DeclareMathSizes{10}{10}{8}{6}
\DeclareSIUnit{\persqrthz}{\ensuremath{\text{\hertz}^{-1/2}}}
\renewcommand{\arraystretch}{1.5}

\titleformat{\section}[runin]{\normalfont\bfseries}{\thesection.}{0.5em}{}[ -]
\titlespacing{\section}{0pc}{5mm}{2mm}

\begin{document}
\onecolumngrid

\title{Numerical Modeling and Experimental Demonstration of Pulsed Charge Control for the Space Inertial Sensor used in LISA}
\makeatletter


\makeatother





\def\addressaa{Universit\'e de Paris, CNRS, Astroparticule et Cosmologie, F-75013 Paris, France}
\def\addresscc{Universit\'e de Strasbourg, CNRS, ISIS, 8 alle\'e Gaspord Monge, F-67000, Strasbourg, France}
\def\addressbb{Department of Mechanical and Aerospace Engineering, MAE-A, P.O. Box~116250, University of Florida, Gainesville, Florida 32611, USA}
\def\addressdd{Department of Physics, 2001 Museum Rd, University of Florida, Gainesville, Florida 32611, USA}
\author{H.~Inchausp\'e}\altaffiliation[Current Address: ]{\addressaa}\affiliation{\addressbb}
\author{T.~Olatunde}\affiliation{\addressbb}
\author{S.~Apple}\affiliation{\addressbb}
\author{S.~Parry}\affiliation{\addressbb}
\author{B.~Letson}\affiliation{\addressbb}
\author{N.~Turetta}\altaffiliation[Current Address: ]{\addresscc}\affiliation{\addressbb}
\author{G.~Mueller}\affiliation{\addressdd}
\author{P.~J.~Wass}\affiliation{\addressbb}
\author{J.~W.~Conklin}\affiliation{\addressbb}
\date{\today}

\graphicspath{{Images/}}

\begin{abstract}
\noindent Electrostatic charge control of isolated free-falling test masses is a key enabling technology for space-based gravitational missions. 
Contact-free electrostatic charge control can be achieved using photoelectron emission from metal surfaces under illumination with deep UV light. A contact-free method minimizes force disturbances that can perturb measurements or interrupt science operations.
In this paper we present charge control experiments using a gravitational reference sensor geometry relevant to the Laser Interferometer Space Antenna (LISA) gravitational wave observatory in a torsion pendulum apparatus. We use a UV LED light source to control the test mass potential, taking advantage of their high bandwidth to phase-lock the light output to 100\,kHz electric fields used for capacitive position sensing of the test mass. We demonstrate charge-rate and test mass potential control by adjustment of the phase of the light with respect to the electric field.
We present a simple physics-based model of the discharging process which explains our experimental results in terms of the UV light distribution in the sensor, surface work functions and quantum yields. A robust fitting method is used to determine the best-fit physical parameters of the model that describe the system. 

\end{abstract}

\newcommand{\indice}[1]{{\scriptscriptstyle #1}}
\newcommand{\exposant}[1]{{\scriptscriptstyle #1}}
\newcommand{\myvec}[2]{\vec{#1}_\indice{#2}}
\newcommand{\myexpr}[3]{#1_\indice{#2}^\exposant{#3}}
\DeclarePairedDelimiterX{\norm}[1]{\lVert}{\rVert}{#1}
\newcommand{\ddt}[2]{ \frac{d}{dt} \left( #1 \right)_#2}
\newcommand{\ddtddt}[2]{ \frac{d^2}{dt^2} \left( #1 \right)_#2}
\newcommand{\myhyperref}[1]{\hyperref[#1]{\ref{#1}}}
\newcommand{\identite}[1]{{\displaystyle \mathbb{1}_{\indice{#1}}}}
\newcommand{\myskew}[2][]{\@ifmtarg{#1}{\left[ #2 \right]^{\times}}{\left[ #2 \right]^{\times, #1}}}
\newcommand{\myat}[2][]{#1|_{#2}}
\newcommand{\timestentothe}[1]{\times 10^{#1}}
\renewcommand{\arraystretch}{1.5}

\maketitle

\section{Introduction}
\label{section: Introduction}
Sensitive space-based gravitational experiments make use of free-falling test masses as inertial references used in tests of fundamental physics \cite{Everitt2015, Touboul2017, Sumner}, studies of the Earth's geopotential \cite{Drinkwater2003, Kornfeld2019} and gravitational wave observatories  \cite{LISA2017, Kawamura2019, Luo2016, Hu2016}. The most sensitive of these make use of completely mechanically and electrically isolated test masses requiring a non-contact method to maintain a neutral net electrostatic charge. In this paper, we present a study of test-mass charge control using photoemission produced by pulsed UV light
in an experimental set up relevant to  Laser Interferometer Space Antenna (LISA) gravitational wave observatory instrumentation.

LISA is a European Space Agency mission being developed in partnership with NASA \cite{LISA2017}. It will be the first space-based gravitational wave observatory, with a planned launch in the 2030s. The observatory will target the low-frequency gravitational-wave frequency band between 100\,$\mu$Hz and 1\,Hz at a strain sensitivity of $10^{-20} /\sqrt{\mathrm{Hz}}$. To achieve this goal, a constellation of three spacecraft shall be used to perform laser interferometry between free-falling test masses over a 2.5\,million-km baseline at 10\,pm$/\sqrt{\mathrm{Hz}}$ precision. The test masses, 46\,mm, 1.96\,kg cubes of gold-platinum alloy, are shielded from the disturbances of the space environment by their spacecraft. They are enclosed by a gravitational reference sensor (GRS) that enables the test mass to maintain a pure free-fall or drag-free condition by providing a low-force-noise environment and six degree-of-freedom sensing and actuation. The residual acceleration of the test masses in the LISA band must be of order fm\,s$^{-2}/\sqrt{\textrm{Hz}}$. The LISA Pathfinder mission \cite{McNamara2008} demonstrated the functionality and performance of a GRS at the level required for LISA \cite{Armano2016, Armano2018c}.

\subsection{Gravitational Reference Sensor}
\label{subsection: Gravitational Reference Sensor}

\begin{figure}
\begin{center}
\includegraphics[scale=0.3]{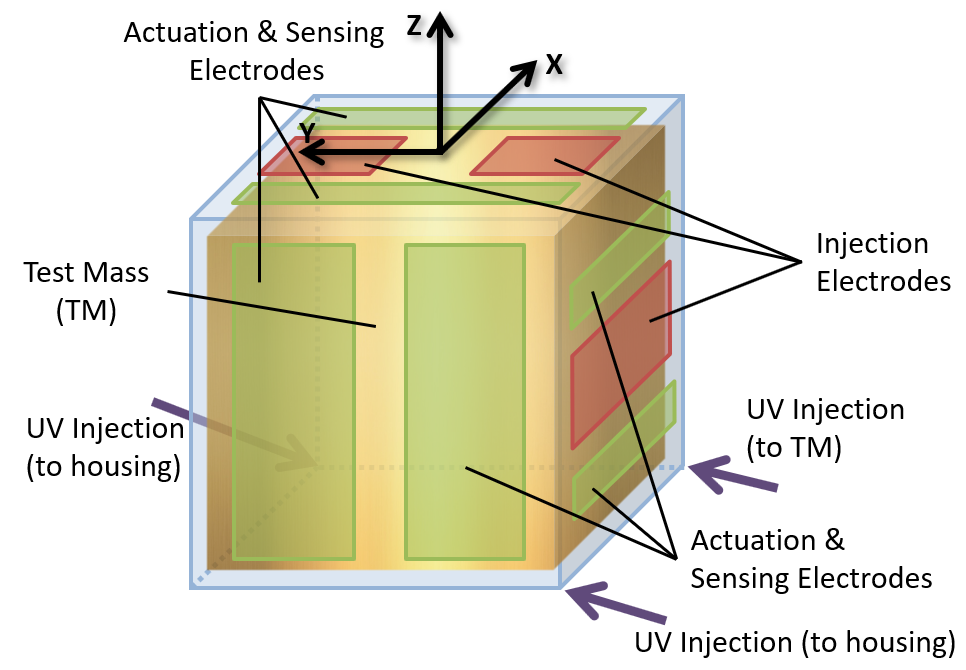}
\end{center}
\caption{Drawing of LPF GRS. Note that the electrode orientation is the same on each electric housing opposing surface.}
\label{figure: LPF GRS}
\end{figure}

The Gravitational Reference Sensor (GRS), performs capacitive sensing and electrostatic actuation of the test mass in six degrees-of-freedom \cite{Dolesi2003}. This, together with an interferometric readout of the test mass position along the line-of-sight of the interferometer arms, allows the spacecraft to follow the test mass motion using micro-Newton thrusters in certain axes \cite{Armano2019}, while applying nano-Newton forces to the test mass to maintain its position inside the GRS. The LISA GRS (Figure \ref{figure: LPF GRS}) is made up of the test mass (TM), surrounded by six pairs of sensing and actuation electrodes and three additional pairs of electrodes responsible for capacitively inducing a 100\,kHz potential on the test mass for position and attitude readout. The GRS also comprises the front-end electronics (FEE) responsible for detecting capacitive changes produced by test mass motion \cite{Armano2017b} and generating actuation signals, launch-lock and precision release mechanisms for the test mass and a system of contactless electrostatic discharge to maintain the test mass at a neutral potential \cite{Armano2018a}. The GRS, coupling with environmental factors defines the force noise environment of the test mass 
\cite{Dolesi2003}.

\subsection{Charge Management}
\label{subsection: Charge Management}

Electrostatic forces acting on the test mass arising from stray potentials and the test mass charge contribute to the acceleration noise budget \cite{Sumner2020}. Energetic charged particles from cosmic rays and solar energetic particles can deposit charge on the TM \cite{Wass2005, Araujo2005} either by stopping directly or by creating secondaries in the spacecraft structure. The leading sources of charge-related noise arise from the charge coupling with noisy electric fields within the GRS and fluctuations in the charge rate from cosmic rays which couple to DC fields \cite{Antonucci, Armano2017a}. In order to mitigate the former effect, and maintain the charge-related acceleration noise within the level allocated in the performance budget, the absolute charge must be held below 15 million elementary charges (2.4\,pC) \cite{Armano2018a}.

A non-contact method is used for discharging to minimize force disturbances on the test mass. Electrons are liberated from the gold-coated test mass or surrounding electrodes using the photoelectric effect under illumination with ultra-violet (UV) light and transferred between the test mass and its surroundings. It is well known that gold surfaces exposed to air develop a strongly-bonded and long-lived contamination that lowers the work function from a value of 5.2\,eV to around 4.2 -- 4.7\,eV \cite{Kahn2016} allowing the use of UV light sources in the wavelength range around 250\,nm (5.0\,eV) for charge control.

Charge control using UV light was first demonstrated in Gravity Probe B \cite{Buchman2000, Everitt2015}, and was used successfully in a LISA-like GRS on-board LISA Pathfinder (LPF) \cite{Armano2018a}. In both missions, low-pressure mercury vapor lamps were used to generate UV light for discharging at a wavelength of 254\,nm. In LISA Pathfinder, the test mass could be charged positively or negatively by preferentially illuminating the test mass or the surrounding electrodes. This charge transfer was supported by DC electric fields in the sensor to favor the exchange of charges in the preferred direction.

The LISA discharge system will make use of recent developments in UV light-emitting diode (LED) technology. UV LEDs offer advantages over discharge lamps in terms of size, electrical power efficiency, reduced complexity in the drive electronics, longer lifetime, lower sensitivity to temperature, and most importantly, a high bandwidth. This latter point allows the possibility to operate the LEDs in a pulsed mode, synchronizing the UV illumination with AC electric fields in the GRS and to modulate the time-averaged UV power with a very high dynamic range. Operating in a DC, constant current mode is also possible and allows for the LEDs to be used in the same way as UV lamps were used in LISA Pathfinder.  

\begin{figure}
\begin{center}
\includegraphics[scale=0.4]{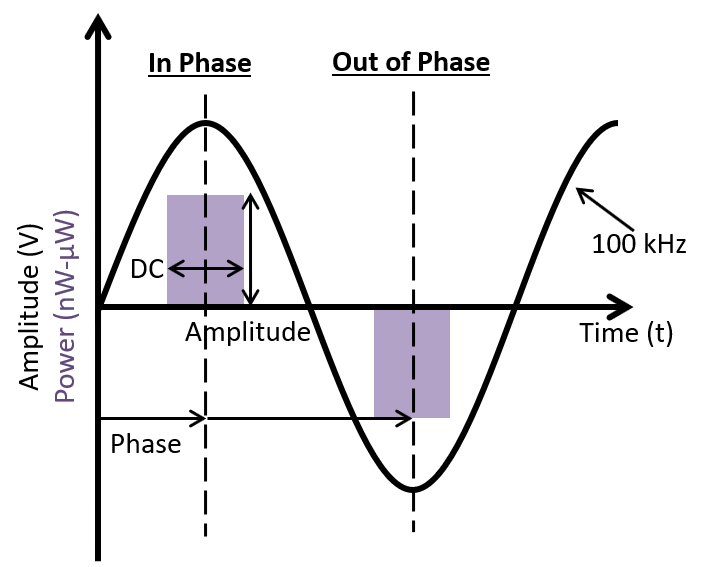}
\end{center}
\caption{Illustration of synchronizing with the 100\,kHz sensing signal for pulsed charge control. Adjusting duty cycle, phase, and amplitude of the UV light is possible. Note that pulsing either in phase or out of phase determines the potential barrier $\Delta V$ that the photoelectrons need to overcome (see section~\ref{section: Model}).}
\label{figure: pulse sync}
\end{figure}

The concept of synchronized, pulsed discharging is well established \cite{Sun2006, Ziegler2014, Olatunde2015} and a number of studies on LED components \cite{Sun2009,Hollington2015, Hollington2017} have demonstrated their suitability for operation in a discharging system for LISA, including a satellite test in low-Earth orbit \cite{Saraf2016a}. 
When operated in a pulsed mode, synchronized with the 100\,kHz capacitive sensing bias in the GRS, the time-averaged power of the light is determined by the peak pulse power, duty cycle and number of pulses per unit time. The phase of the UV pulse relative to the 100\,kHz sensing signal can also be adjusted as indicated in Figure~\ref{figure: pulse sync}.

The small gaps and reflective surfaces of the GRS (see for example Figure \ref{figure: LPF GRS} and \ref{fig:AC port}), mean that multiple reflections of UV light between the test mass and electrodes are inevitable. The result is that the net test mass discharging current is made up of opposing electron flows. The balance between the flows depends on the potential drop between the two surfaces (in a way that is discussed further in section \ref{section: Model}) so that the two flows can be balanced and an equilibrium reached. By controlling the phase of the UV illumination relative to the AC electric field in the GRS, both the charge rate of a neutral test mass and the equilibrium test mass potential can be controlled.

In this paper we present for the first time the results of synchronized charge control in a LISA-like GRS geometry used to control the discharge rate and test mass potential together with a simple theoretical model of the discharging process. The paper is structured as follows: first, we present a discussion of the theoretical charge control model that describes the discharging processes inside the GRS. In section \ref{section: Torsion Pendulum Measurements}, we present a set of experimental charge control measurements using pulsed UV light from UV LEDs obtained with a torsion pendulum equipped with a representative GRS. We discuss the results of these measurements and analyze them in the context of the discharge model. Finally we discuss how well our model describes the experimental data and present our conclusions.

\section{Charge control model based on electron exchange between two parallel plates}
\label{section: Model}

\begin{figure}
\label{fig:1}
\begin{center}
\includegraphics[scale=0.7]{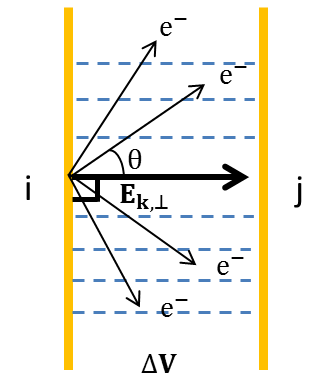}
\end{center}
\caption{Illustration of photoelectrons emitted from a plane surface in a parallel plate configuration showing electron velocities and electric field lines.}
\label{fig: ElEnDist}
\end{figure}

\begin{figure}
\begin{center}
\includegraphics[width=0.483\textwidth]{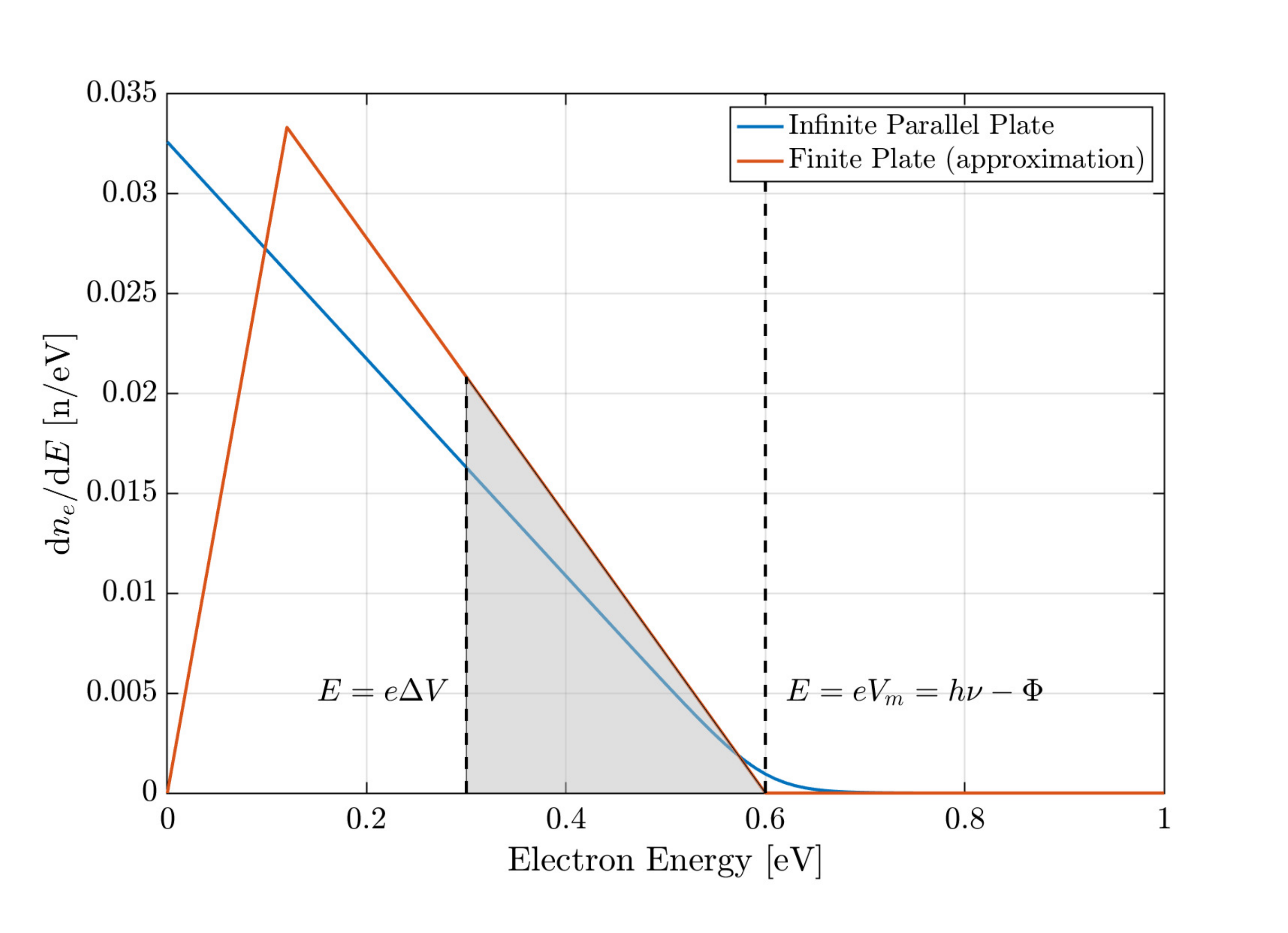}
\end{center}
\caption{Energy distribution of electrons derived from the component of the electron velocity normal to the emitting surface. Blue curve indicates the distribution derived in \cite{Hechenblaikner2012} for an infinite parallel plate geometry. The red curve is the distribution used in our model and by \cite{Armano2018a} incorporating the effects of a 3-D sensor geometry. Also indicated are the minimum energy of electrons able to traverse the gap, $e \Delta V$ and the maximum energy of electrons $h\nu-\Phi$.}
\label{fig: electDistSimple}
\end{figure}

In the GRS, each TM surface and the opposing electrode in the electrode housing form a simple parallel plate capacitor. The successful extraction and transfer of an electron from one surface to another against an opposing voltage $\Delta V$ depends on the kinetic energy perpendicular to the plates as sketched in Figure \ref{fig: ElEnDist}.
Electrons are assumed to be emitted from the surface with a cosine angular distribution. In the case of interest here in which the energy, $h\nu$, of a photon with optical frequency $\nu$ is comparable to the work function of the illuminated metal $\Phi$, the energy distribution of emitted electrons is approximated by a near-linear function.
In \cite{Hechenblaikner2012} the energy distribution of electrons able to traverse a gap between two surfaces with a potential barrier $\Delta V$ was derived for a number of geometries, ranging from a point source surrounded by a sphere to the parallel plate case described above. In this latter case, the distribution taking into account thermal energy of electrons in the metal and tunneling effects is described by 
\begin{center}
\begin{equation}\label{eq: electDistFull}
f(\Delta V) \propto \ln (\exp{\frac{(-\Delta V+V_m)e}{k_B T}+1)},
\end{equation}
\end{center}
where $e$ is the charge of an electron, $k_B$ is Boltzmann's constant, $T$ is the temperature of the metal and $e V_m= h\nu-\Phi$ is the maximum kinetic energy an electron can have after being extracted by a photon from the surface. The relationship in Equation \ref{eq: electDistFull} is represented by the blue curve in Figure \ref{fig: electDistSimple}, showing that, with the exception of the region around the maximum electron kinetic energy, $eV_m$, the function is well approximated by a straight line with a negative slope in the region $0<\Delta V<V_m$ , and zero for $\Delta V > V_m$
\begin{equation}
f(\Delta V) \propto \frac{-e}{k_BT}(\Delta V - V_m).
\label{1}
\end{equation}

In a GRS system such as that shown in Figure \ref{figure: LPF GRS}, a parallel plate model is not fully valid; on the TM faces, a parallel plate is a good approximation but close to the test mass edges and corners the geometry can resemble a point source emission within a sphere. As a result the net energy distribution of transferred electrons will take a triangular form shown in Figure \ref{fig: electDistSimple}. This shape has been experimentally verified in laboratory experiments \cite{Wass2019}, and has been found to be compatible with measurements made in a LISA-like GRS in the LISA Pathfinder mission \cite{Armano2018a}. 

The distribution is parameterized in the following way: if the slope of the red curve, $\frac{e}{k_B T}$ in Figure \ref{fig: electDistSimple} is defined as $\beta_0$ and the positive and negative slopes in the blue curve are $\beta_1$ and $\beta_2$ respectively, the total energy distribution for Figure \ref{fig: electDistSimple} can be written in terms of $\beta_0$ in the Equation below:
\begin{equation}
  f(\Delta V)=\begin{cases}
    \frac{\beta_0}{x}\Delta V, & \text{if $\Delta V < xV_m$},\\
    \frac{\beta_0}{1-x} (V_m -\Delta V), & \text{if $\Delta V > xV_m$},\\
        0, & \text{if $\Delta V > V_m$}.\\
  \end{cases}
\label{eq: ElEnDist}
\end{equation}
The shape of the triangular distribution is defined by the position of the summit, $x$, expressed as a fraction of $V_m$, has been measured to be about 0.2 in the case of LPF geometry \cite{Wass2019,Armano2018a}. This value will be used in the present analysis.

The electron flow rate from surface $i$ to surface $j$, $\dot{n}_{ij}$ is expressed in Equation \ref{eq: ndot} in terms of the following; the time-averaged pulsed UV power $P_{\rm{UV}}$ injected into the GRS (expressed as photon flux), the fraction of the total power absorbed by the $i$th surface $\alpha_i$, the Quantum Yield $QY_i$ of the surface, the distribution of  photon energies emitted by the light source $g(\nu)$, defined such that $h\int \nu g(\nu)d\nu=P_{\rm{UV}}$ and $W_i$, defined in Equation \ref{eq: emissionRatio} as the fraction of photoelectrons emitted from surface $i$ with sufficient energy to overcome the potential barrier $\Delta V_{ij}$ between surface $i$ and $j$. 

\begin{center}
\begin{equation}\label{eq: ndot}
\dot{n}_{ij} = \alpha_i\ QY_i\ W_i \frac{\ P_{UV}}{ h \int_{0}^{+\infty}\nu g_i(\nu) d\nu} \int_{\frac{\phi_i}{h}}^{+\infty} g_i(\nu) d\nu
\end{equation}
\label{eq: Electron flow rate}
\end{center}

\begin{center}
\begin{equation}\label{eq: emissionRatio}
\ W_i = \frac{\int_{\frac{e\Delta V_{ij} + \phi_i}{h}}^{+\infty} g_i(\nu) d\nu \int_{\Delta V_{ij}}^{V_{mi}} f_i(\phi_i,\Delta V_{ij},\nu,T) d\Delta V_{ij}}{\int_{\frac{\phi_i}{h}}^{+\infty} g_i(\nu) d\nu\int_{0}^{V_{mi}} f_i(\phi_i,\Delta V_{ij},\nu,T) d\Delta V_{ij}}
\end{equation}
\end{center}

The ratio of the integrals over $g(\nu)$ in Equation \ref{eq: ndot} accounts for the fraction of $P_{\rm{UV}}$ contributed by photons with energy below the work function. If all photons have an energy larger than the work function then this simplifies to $\alpha_i QY_i P_{\rm{UV}}$.  $W_i$ integrates over the photon and electron energy distributions with sufficient energy to overcome the potential barrier $\Delta V$. $W_i = 1$ in the case where $\Delta V_{ij}<0$.

For a UV LED expected to be used in the LISA charge management system, the energy distribution of the photons, $g(\nu)$ can be approximated by a Gaussian function \cite{Hollington2015}. The minimum energy a UV photon must have in order to generate photo electrons that can contribute to $\dot{n}_{ij}$ is $\left|\frac{e\Delta V + \phi}{h}\right|$ which forms the lower limit of the $g(\nu)$ integral in Equation (\ref{eq: ndot}). The resolution of the double integral assuming the Gaussian distribution for photons and the energy distribution for photo electrons shown in Figure \ref{fig: electDistSimple} can be found in Appendix 1.

\subsection{The GRS as parallel plate capacitors}

Adopting the same approach as \cite{Armano2018a}, the LISA GRS, shown in Figure \ref{figure: LPF GRS}, can be thought of as a system of parallel plate capacitors responsible for AC capacitive biasing, position sensing and actuation of the test mass. In the experimental set up we describe in the following section, we make use of a simplified sensor (illustrated in Figure \ref{fig:UV LED geometries}) with only three pairs of electrodes, one per axis. One axis is used for capacitively injecting a 100\,kHz potential on the test mass for capacitive sensing, while the other two are used for position sensing. Unlike the LISA GRS, AC actuation signals are not used. 

\begin{figure}
\includegraphics[scale=0.54, trim={10.0cm 5.0cm 9.8cm 6.0cm}, clip]{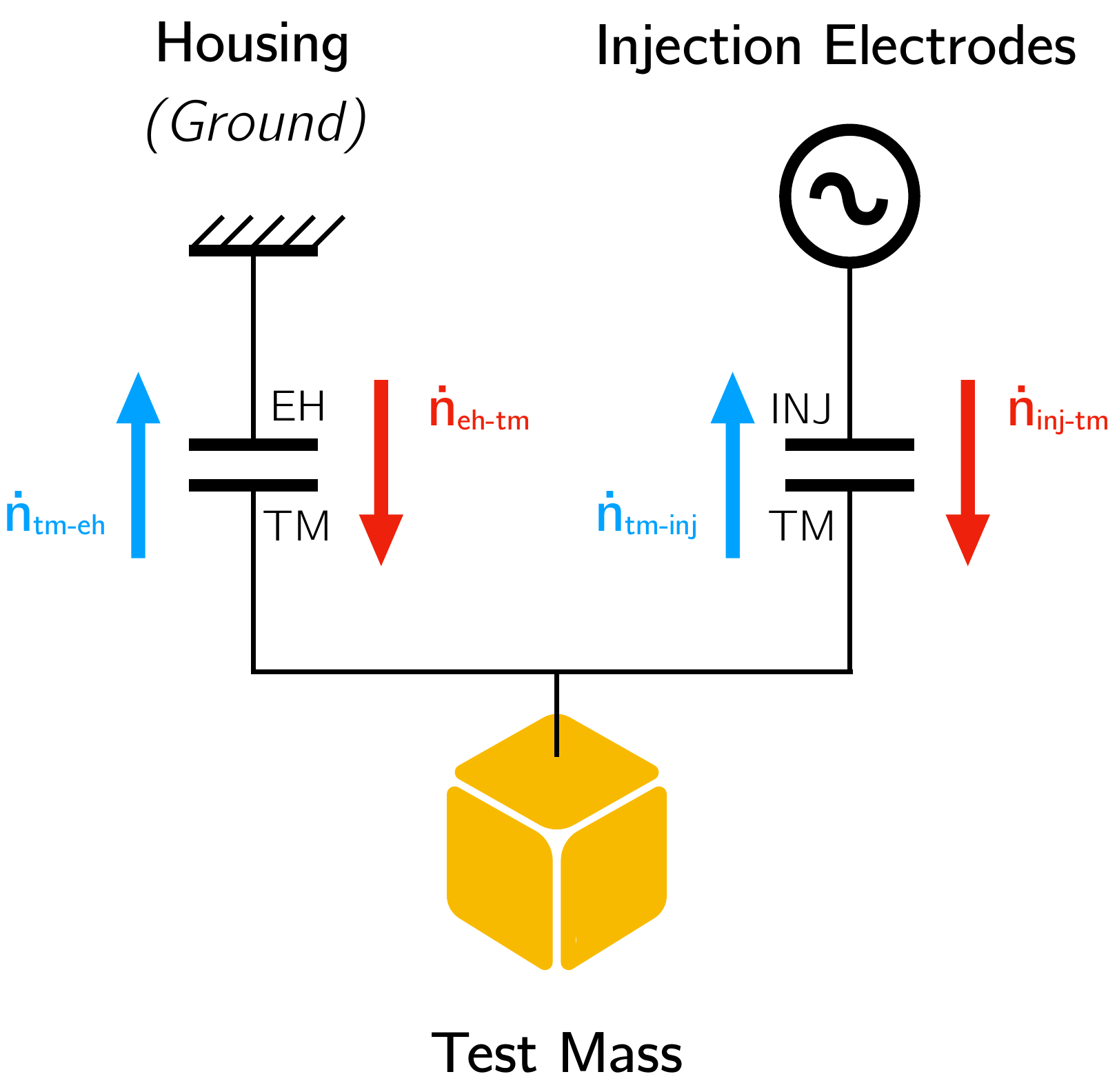}
\caption{Simplified GRS equivalent circuit for consideration of electric fields relevant to discharging.}
\label{fig: CircuitModel}
\end{figure}

In the absence of actuation signals, and ignoring small changes in capacitance due to test-mass motion, the model of parallel plate capacitors can be simplified further to a system of two capacitors in series as shown in Figure \ref{fig: CircuitModel}. The capacitor plates represent the surfaces illuminated by the UV radiation in the GRS that are at a common instantaneous potential. One capacitor represents the `injection' electrodes and test mass surfaces opposite (hereafter labeled inj-tm and tm-inj), the other the combination of sensing electrodes and electrode housing surfaces held at zero potential and the test mass surfaces opposite (eh-tm and tm-eh).   


 The capacitance of the electrodes towards the TM and the capacitance of the TM toward ground are respectively denoted as $C_{\rm{inj}}$ and $C_{\rm{T}}$ while the injection voltage is represented as $V_{\rm{inj}}$. 
 The instantaneous potential of the test mass $V_{\rm{TM\,(inst)}}$ with respect to the grounded electrode housing has two contributions. The first comes from the net charge $q$ on the entire test mass and the second is 
 an induced potential coming from voltages applied to the electrodes. In the particular case of the injection voltage, we define $\eta$ as the ratio of the combined capacitance of the injection electrodes to the capacitance of the test mass such that
\begin{align}
V_{\rm{TM\,inst}} &= \frac{q}{C_{\rm{T}}}+ \frac{\sum_{i}{C_{i}}{V_{\rm{inj}}}}{C_T} \nonumber \\
&= V_{\rm{TM}} + \eta V_{\rm{inj}}.
\label{eq: net voltage on TM}
\end{align}
The DC test mass potential $V_{\rm{TM}}$ is the relevant parameter in determining test mass force disturbances and the observable in our torsion pendulum measurements.  The rate of change of test mass potential under UV illumination can be expressed in terms of the photoelectron flows, $\dot{n}$ between each of the relevant surfaces in the GRS:
\begin{center}
\begin{equation}\label{eq: Net potential rate}
\frac{\mathrm{d}V_{TM}}{\mathrm{d}t} =  \frac{e}{C_T}(\Dot{n}_{\rm{inj-tm}}-\Dot{n}_{\rm{tm-inj}}+\Dot{n}_{\rm{eh-tm}}-\Dot{n}_{\rm{tm-eh}}).
\end{equation}
\end{center}
Each $\dot{n}$ term in Equation (\ref{eq: Net potential rate}) depends on the potential drop between the surfaces, varying from zero when the potential drop is greater than $h\nu - \Phi$ to maximum when the potential drop is less than or equal to zero. The potential drop between test mass and injection electrode $\Delta V_{\rm{inj-tm}}$ is given by the difference between the electrode voltage at the instant of illumination, $V_{\rm{inj}}$ and the test mass voltage $V_{\rm{TM}}$, made up of the DC potential of the test mass from charge accrual and the instantaneous capacitive polarization by $V_{\rm{inj}}$ through $\eta$. Each term has a corresponding illumination fraction $\alpha$ which we combine in vectorial form $\vec{\alpha} = [\alpha_{\mathrm{inj-tm}},\, \alpha_{\mathrm{tm-inj}},\,\alpha_{\mathrm{eh-tm}},\, \alpha_{\mathrm{tm-eh}}]$.
The maximum photocurrent that can be transferred from one surface to another is given by Equation (\ref{eq: ndot}). If the QY of each surface is assumed equal, since $P_{\rm{UV}}$ is also constant, the photoelectron current is determined by the fraction of light illuminating that surface. 
\begin{figure}
\includegraphics[width=0.483\textwidth]{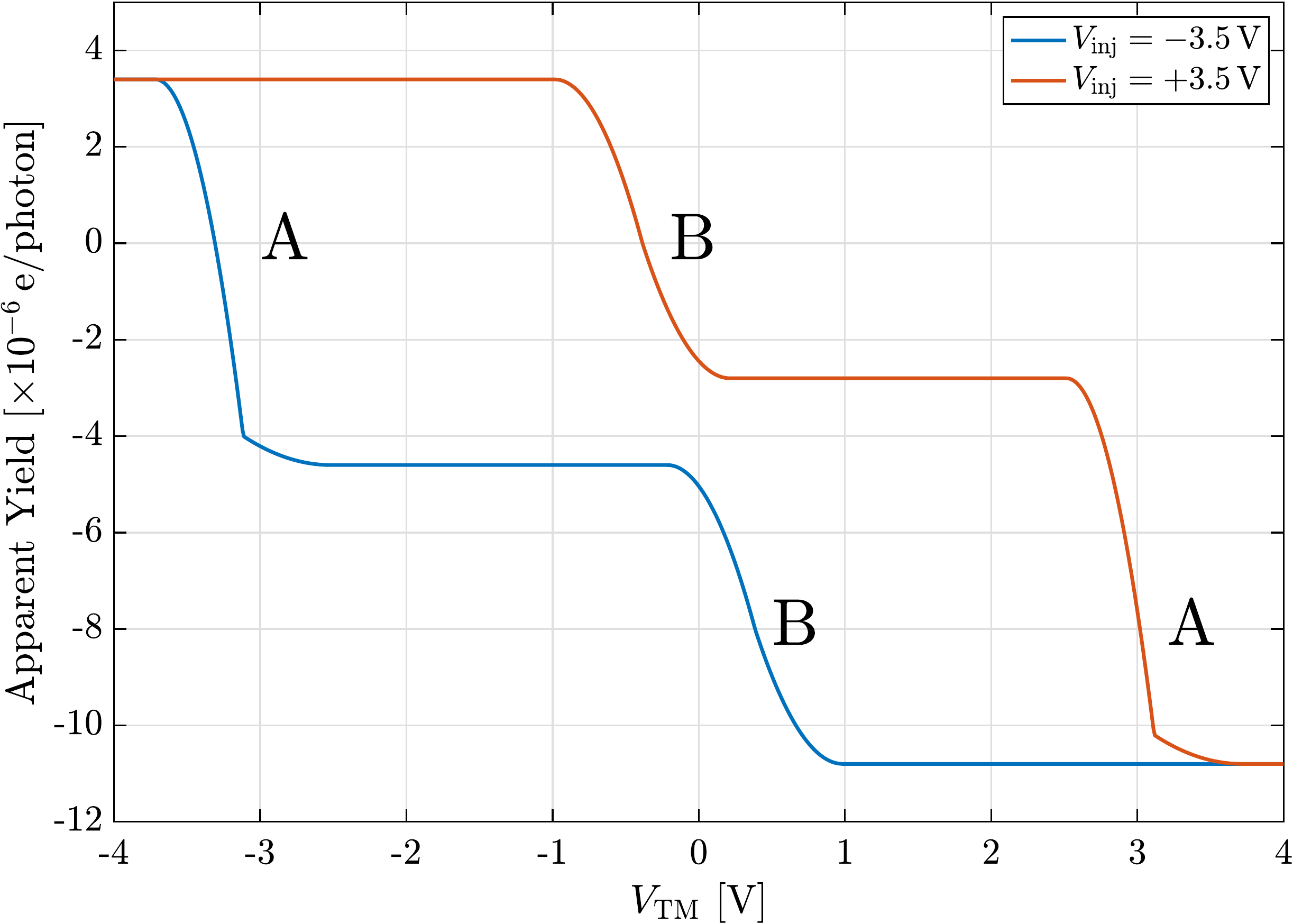}
\caption{Simple model prediction of apparent yield (charges transferred from the test mass per photon injected into the system) as a function of DC test mass potential for positive and negative values of $V_{\rm{inj}}$. Transition regions `A' occur when the test mass DC potential approaches the injection potential. Transition regions `B' occur around $-\eta V_{\rm{inj}}$}
\label{fig:transition_regions}
\end{figure}

In understanding the behavior of the integrated discharging system it is informative to consider the charging rate as a function of the DC test mass potential for a given $V_{\rm{inj}}$. The charge rate behavior expressed in terms of apparent yield as a function of $V_{\rm{TM}}$ for a candidate sensor with assumed properties is provided in Figure \ref{fig:transition_regions}. Apparent yield is defined as the net discharge current per photon injected into the system. The curves are calculated assuming a monochromatic light source with a photon energy of 5.0\,eV illuminating surfaces with a work function $\Phi=4.4$\,eV and a simple triangular energy distribution for electrons, as shown in Figure \ref{fig: ElEnDist} but with a peak at 0\,eV. The ratio of capacitances $\eta=0.11$ and the assumed illumination favors negative test mass charging with $\vec{\alpha}=[0.37,\,0.03,\,0.17,\,0.14]$. The curves show ranges of constant charge rate, separated by two transition regions which correspond to the test mass potential ranges in which the net current in each of the two capacitors in the model change direction. The width of the transition is determined by $h\nu_{\rm{max}}-\Phi$. Transition A, centered on $V_{\rm{TM}} = V_{\rm{inj}} - \eta V_{\rm{inj}}$ is associated with the inj-tm photocurrent flow. Transition B is centered on $V_{\rm{TM}} = - \eta V_{\rm{inj}}$ and is associated with the eh-tm photocurrent flow. The plateau levels on the curve are set by the products $\alpha_i QY_i$, corresponding to the combination of saturation photocurrents in the two capacitors. The two curves calculated at $V_{\rm{inj}}=\pm3.5$\,V illustrate the relative shifts in the values of $V_{\rm{TM}}$ associated the two transition regions.
\begin{figure}
\includegraphics[width=0.483\textwidth]{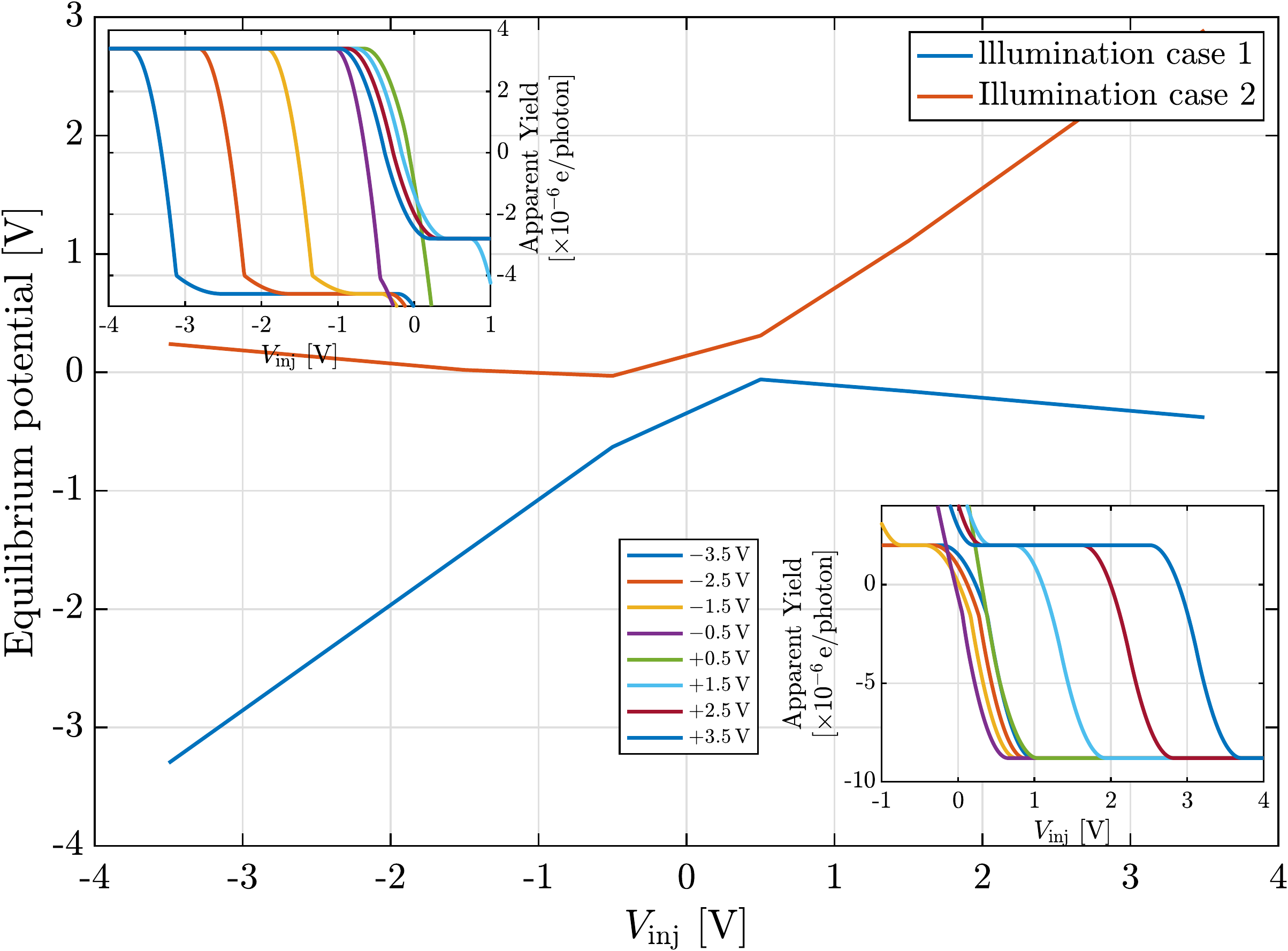}
\caption{Simple model prediction of equilibrium potentials for two different illumination cases. Equilibrium potentials are extracted from the zero-crossing of apparent yield curves such as those shown in Figure \ref{fig:transition_regions}. Insets show a range of such curves from $V_{\rm{inj}}=-3.5$-$+3.5$\,V for illumination case 1 (negative charging), top-left and case 2 (positive charging), bottom-right highlighting the zero-crossing region. The legend shown refers to both inset graphs.}
\label{fig: EqPotential_SimpleModel}
\end{figure}
The point at which the curve crosses the apparent yield axis (where $\mathrm{d}V_{\rm{TM}}/\mathrm{d}t=0$) is the equilibrium test mass potential and may occur during transition A or B depending on whether the middle plateau is positive or negative. For an equilibrium potential in transition A we require illumination case 1: 
\begin{equation}
    \alpha_{\rm{inj-tm}} QY_{\rm{inj-tm}}>\alpha_{\rm{eh-tm}} QY_{\rm{eh-tm}},
\end{equation}
and for transition B, illumination case 2:
\begin{equation}
    \alpha_{\rm{tm-eh}} QY_{\rm{tm-eh}}>\alpha_{\rm{inj-tm}} QY_{\rm{inj-tm}}.
\end{equation}
Since the aim of certain operational test mass discharge modes is to control the test mass potential to equilibrium we can also examine the dependence of the equilibrium potential on $V_{\rm{inj}}$, the instantaneous injection electrode potential (Figure \ref{fig: EqPotential_SimpleModel}). To a first approximation, the dependence on injection potential of the equilibrium potential is equal to the dependence of the transition position in which the apparent yield crosses zero. This crossing may occur in transition A or B depending on illumination and $V_{\rm{inj}}$. As $V_{\rm{inj}}$ increases from negative towards positive values, the order of the transitions reverses once $V_{\rm{inj}}$ exceeds the potential of transition B. If the equilibrium lies during transition A, (illumination case 1) the equilibrium potential will vary as $V_{\rm{inj}}$ while $V_{\rm{inj}}$ is below transition B and as $-\eta V_{\rm{inj}}$ when $V_{\rm{inj}}$ is greater than $V_{\rm{TM}}$ at transition B. Conversely if the equilibrium lies in transition B, the dependence will be $-\eta V_{\rm{inj}}$ for $V_{\rm{inj}}$ below transition B and $V_{\rm{inj}}$  above. This behaviour is illustrated in Figure \ref{fig: EqPotential_SimpleModel}, where we plot the dependence of the apparent yield zero-crossings 
at a range of values of $V_{\rm{inj}}$ for the illumination scenario modelled in Figure \ref{fig:transition_regions} and a second scenario representative of illumination case 2, favoring positive test mass charging with $\vec{\alpha} = [0.26,\,0.28,\,0.18,\,0.36]$.

In the remainder of the paper we will describe the application of this model to laboratory measurements of test mass discharging using a synchronized scheme in a simplified but representative GRS installed in a low-noise torsion pendulum facility.

\section{Torsion Pendulum Measurements}
\label{section: Torsion Pendulum Measurements}
\begin{figure}
\includegraphics[scale=0.45, frame]{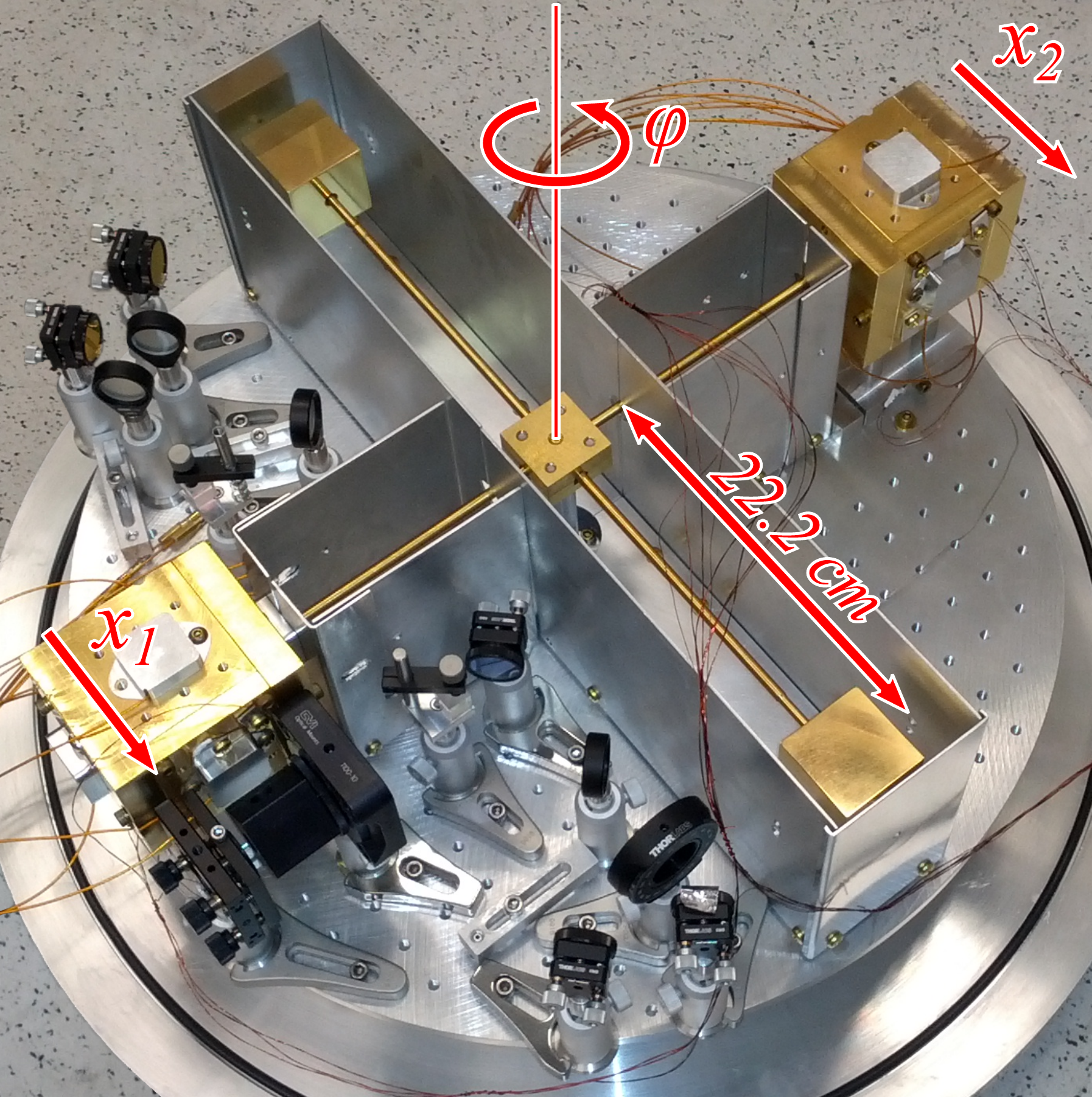}
\caption{View inside the torsion pendulum vacuum chamber with unsuspended cross-bar assembly. Two of test masses are visible while the other two are housed in the respective EHs.}
\label{Crossbar}
\end{figure}

The torsion pendulum facility developed at the University of Florida serves as a test bed for gravitational reference sensors and associated technology for LISA \cite{Ciani2017}. The torsion pendulum consists of a suspended cross-bar assembly with four hollow, gold-coated, aluminum cubic test masses with the same dimensions as the LISA TMs (Figure \ref{Crossbar}. The cross-bar is suspended by a 50\,$\mu$m-diameter, 1\,m-long tungsten fiber. Two of the opposing test masses are enclosed by simplified GRS already described and are electrically isolated from the cross-bar assembly. This enables us to perform test-mass charging and discharging measurements using photoemission in a LISA-like setup. The differential position of the other two test masses is interrogated by a homodyne interferometer, which provides the most sensitive readout of the pendulum angle. 
 
The performance of the pendulum facility can be characterized by assessing the residual torque noise due to surface forces on the TM. Overall performance is limited by a number of factors discussed in detail in \cite{Ciani2017} including the angular readout of the pendulum, surface forces acting on the test masses in the two GRS, environmental disturbances, and the thermal noise of the suspension fiber. 
The typical performance of the pendulum during the study reported here is approximately $3\times 10^{-13}$\,N\,m/$\sqrt{\textrm{Hz}}$  at 2\,mHz comparable to the performance of other similar instruments \cite{Cavalleri2009, Bassan2016a}.

\begin{figure}
\includegraphics[width=0.48\textwidth]{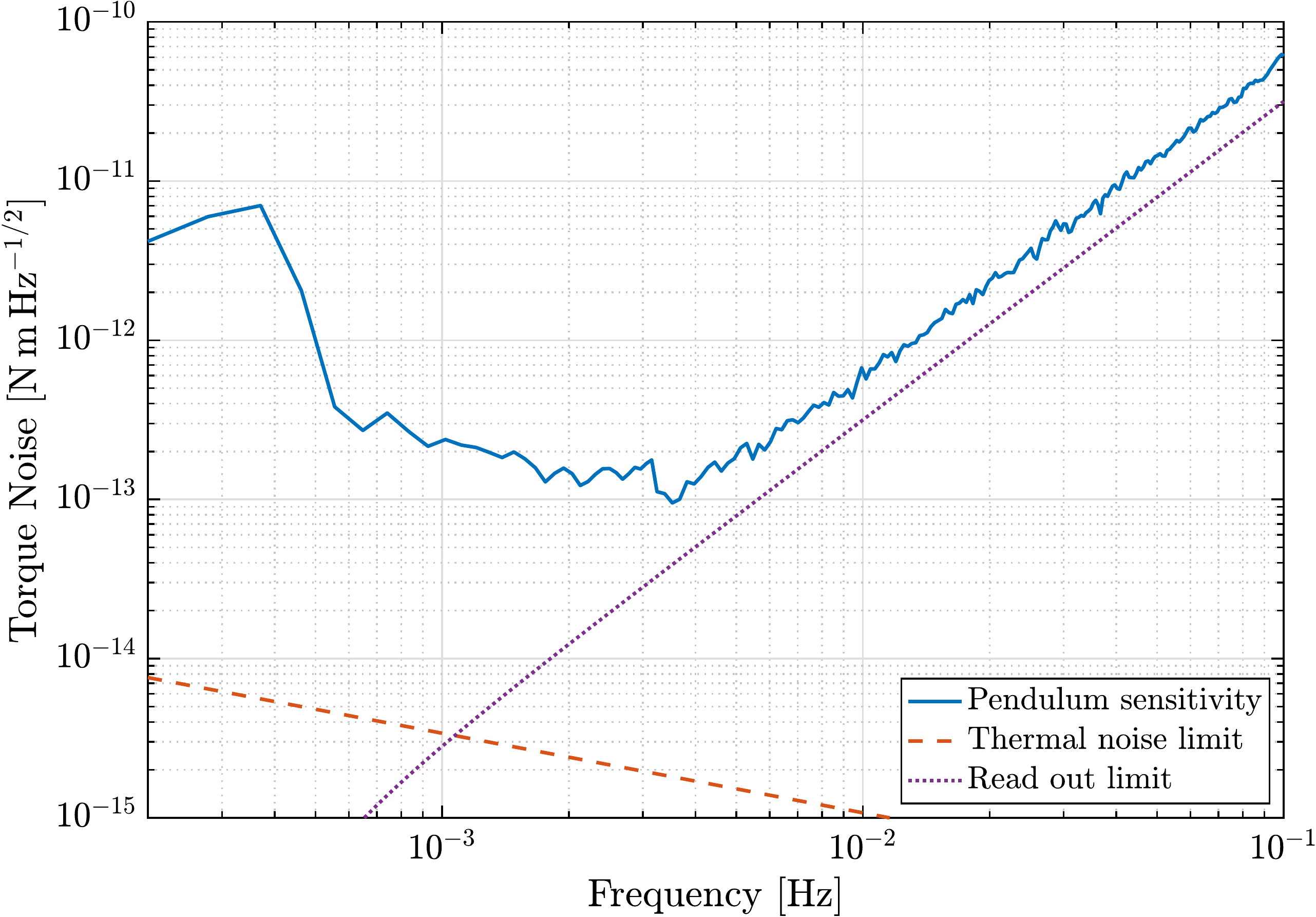}
\caption{Torque sensitivity curve of the UF torsion pendulum from \cite{Ciani2017}}
\label{sensitivity}
\end{figure}

\subsection{Charge Management Experiments}

Measuring charge on the TMs in the torsion pendulum GRS is done by exciting an electrostatic force on the test mass by applying equal and opposite sinusoidal voltages to the pair of electrodes across the $x$ axis of the GRS, tangential to the rotation axis of the pendulum. The coherent torque on the pendulum is proportional to the test mass potential $V_{\mathrm{TM}}$, 
\begin{equation}
    N_{1\omega} \approx rF_x = 2r \left| \frac{\mathrm{d}C_x}{\mathrm{d}x}\right| V_{\rm{TM}}V_{\rm{mod}},
\end{equation}
where $r$ is the pendulum arm, 0.22\,m, $V_{\mathrm{mod}}$ is the excitation voltage applied to the two $x$ electrodes in the GRS, $\frac{\mathrm{d}C_x}{\mathrm{d}x}$ is the derivative with respect to position of the capacitance between the electrode and test mass.
$N_{1\omega}$ is determined from the pendulum angular readout. The measured angle is demodulated at the measurement frequency and converted to torque through the pendulum transfer function providing one measurement point per demodulation cycle. 

The torque sensitivity of the pendulum at the measurement frequency and the amplitude of the modulation sets the charge measurement accuracy. While the optimum sensitivity is at 2\,mHz we chose a faster modulation in order to be able to resolve changes in charge more quickly.

\begin{figure}
    \centering
     \subfloat[][Section of the simplified GRS showing TM and Electrode Housing illumination geometry.]{\includegraphics[scale=0.215, trim={0.0cm 0.0cm 0.0cm 0.0cm}, clip, frame]{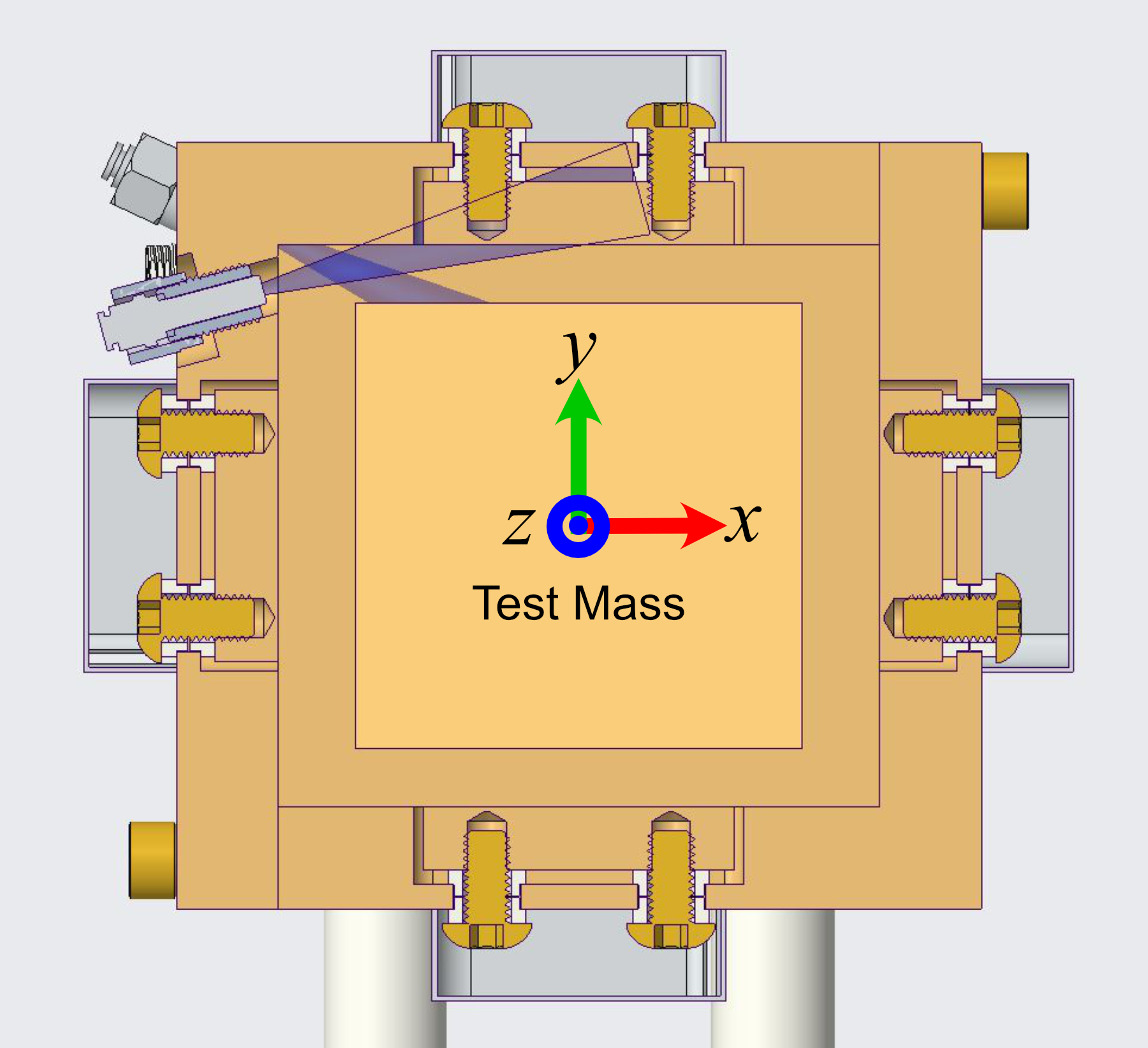}\label{<figure1>}}
     \subfloat[][Photograph of simplified GRS showing injection (bottom face) and position sensing.]{\includegraphics[scale=0.069, trim={2.0cm 10.0cm 8.0cm 8.0cm}, clip, frame]{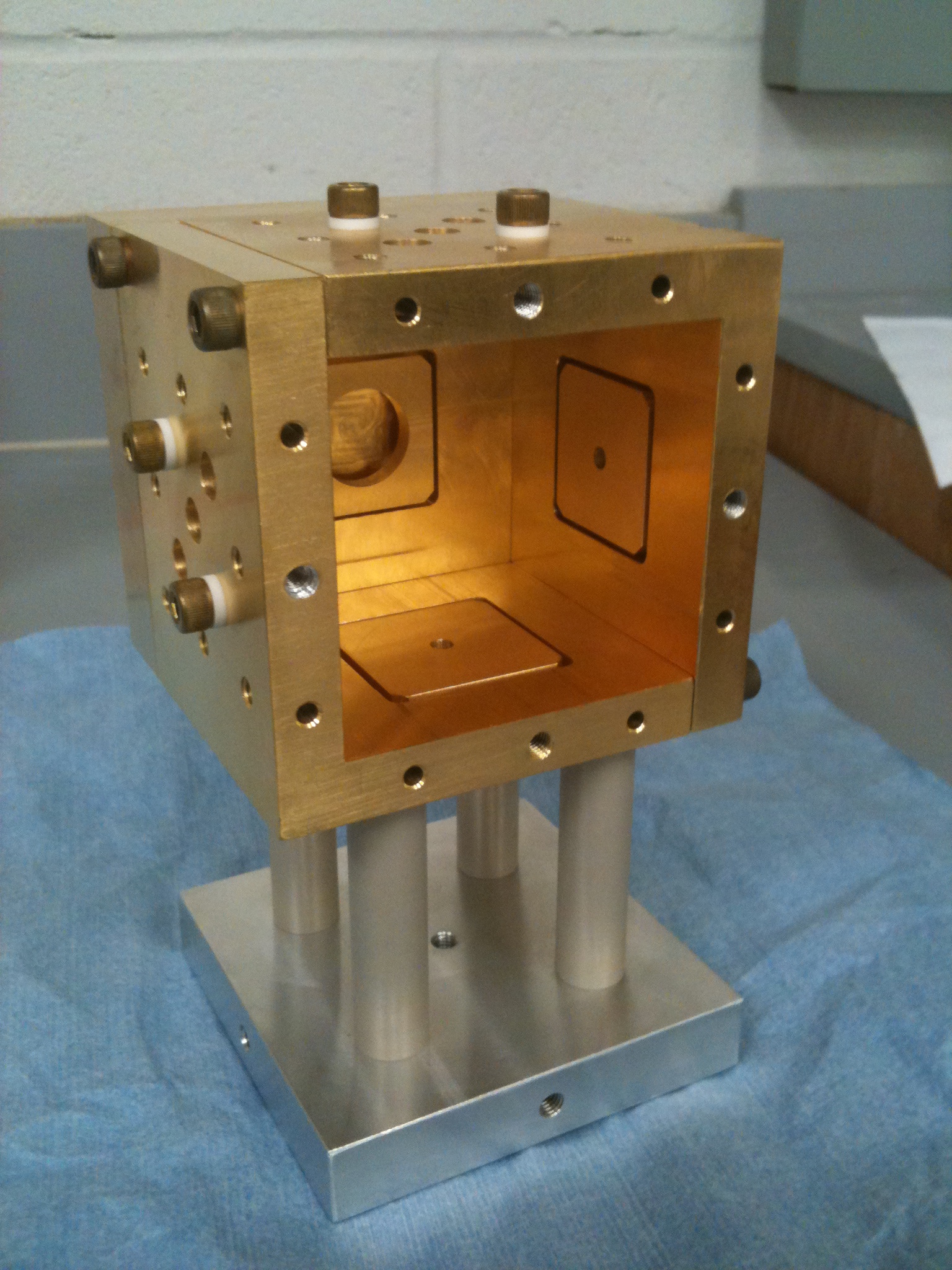}\label{fig:AC port}}
     \caption{Geometry of the simplified GRS.}
     \label{fig:UV LED geometries}
\end{figure}

The GRS in the torsion pendulum has UV fiber optic ports for charge management connected through feedthroughs in the vacuum chamber to the UV LED light source outside the chamber. The net transmission from the input of the vacuum chamber to GRS fiber port has been measured to be between 0.5 and 0.55.
One UV light port is directed toward the EH surface to preferentially transfer electrons onto the test mass, one toward the TM surface to transfer charge in the opposite direction as shown in Figure \ref{fig:UV LED geometries}. The relevant capacitances in our simplified GRS are $C_x = 0.924$\,pF, $\frac{\mathrm{d}C_x}{\mathrm{d}x}=116$\,pF/m, $C_{inj} = 1.60$\,pF and $C_{\mathrm{T}}=20.1$\,pF giving $\eta=0.152$.

The charge control experiments on the pendulum used a prototype UV LED control electronics \cite{Olatunde2015}. This electronics box houses fiber-coupled UV LED with a peak emission at 240\,nm. The electronics drives the UV LED with 100\,kHz pulses synchronized with the capacitive sensing injection signal. The phase, duty cycle and amplitude of the pulses can be set by adjusting analog voltage references.

 A set of experiments were conducted to determine the change in test mass equilibrium potential as a function of the phase offset between the pulsed light and 100\,kHz injection field in the GRS for the two different UV light illumination cases.
 The equilibrium potential was found at 11 different phases ranging from completely in phase ($0\deg$) to completely out of phase ($180\deg$) with the 100\,kHz injection voltage. For each experiment a 5$\%$  duty cycle was used and the phase of the UV light relative to the injection signal was recorded using an oscilloscope. The measurement frequency used was 17\,mHz. Since the test mass equilibrium potential depends on the amplitude of the injection voltage and the electrostatic stiffness associated with the test mass potential that can render the pendulum unstable, the injection voltage used for capacitive sensing was limited to an amplitude of 3.5\,V. Additional limitations on the measurement were imposed by torque spikes thought to be associated with a virtual leak in the system which occur periodically and the range of the interferometer readout limited by surface imperfections on the test masses. The combination of these limitations restrict our ability to make long-duration measurements of $\mathrm{d}V_{TM}/\mathrm{d}t$ used to determine discharge curves such as those presented in \cite{Armano2018a}.

 For each measurement run, the desired phase was set, the fiber-coupled UV LED was connected to the feedthrough corresponding to the desired UV light injection port, and the charge measured continuously until the TM potential reached equilibrium. There were several experiments in which the pendulum motion to became excited out of range of the laser interferometer prior to reaching equilibrium. In these cases, extrapolation using a model fit to the time series data was necessary. 
 
 In the second set of measurements presented here, we determine the charge rate of the TM at $V_{\rm{TM}}=0$\,V in the same measurement conditions as described previously. These experiments are of particular relevance to LISA since the requirement on charge control is to maintain the absolute test mass potential below 70\,mV.  
 In this test, the power out of the fiber coupled UV LED was measured prior to each measurement, and the phase and duty cycle of the pulsed signal driving the UV LED was measured using an oscilloscope. The UV output of the LED was stable throughout all experiments at $150 \pm 10$\,nW.
 At this level of UV power the time for individual charge management runs ranged anywhere from about 10 to 45 minutes depending on the phase of UV light.  
 Charge management runs were begun with a TM potential far enough from 0\,V such that a sufficient number of data points were taken to well characterize the behavior of the TM potential as it approached and passed through 0\,V.

\section{Results}
\label{section: Results}

The two types of experiments performed on the pendulum facility, each studied for two different illumination directions, EH and TM, provide different approaches to probe the reliability of the model. The model parameters influence the measurement results in different ways for each case. Combining all measurements in a global model fit allows for the mitigation of possible parameter degeneracies of the model predictions. This approach provides a rigorous analysis of the system including uncertainties. Previous analyses of similar systems using a similar physical model \cite{Armano2018a}, estimated these parameters in an \emph{ad hoc} manner. In this section, we will present a comparison between our analytical model and $2 \times 2$ sets of 11 measurements obtained using the torsion pendulum. 
We will discuss the fitting methods and strategy used, the best-fit model obtained and how it compares to a simple photon tracking study of the illuminations within the GRS.

\subsection{Data Preprocessing}
\label{subsection: Data Preprocessing}

\begin{figure*}
    \centering
    \subfloat[][EH port, $\phi = 54 \si{\degree}, V_{\rm{inj}}=2.83$\,V]{\includegraphics[scale=0.45, trim={4.5cm 1.0cm 4.5cm 0.0cm}, clip]{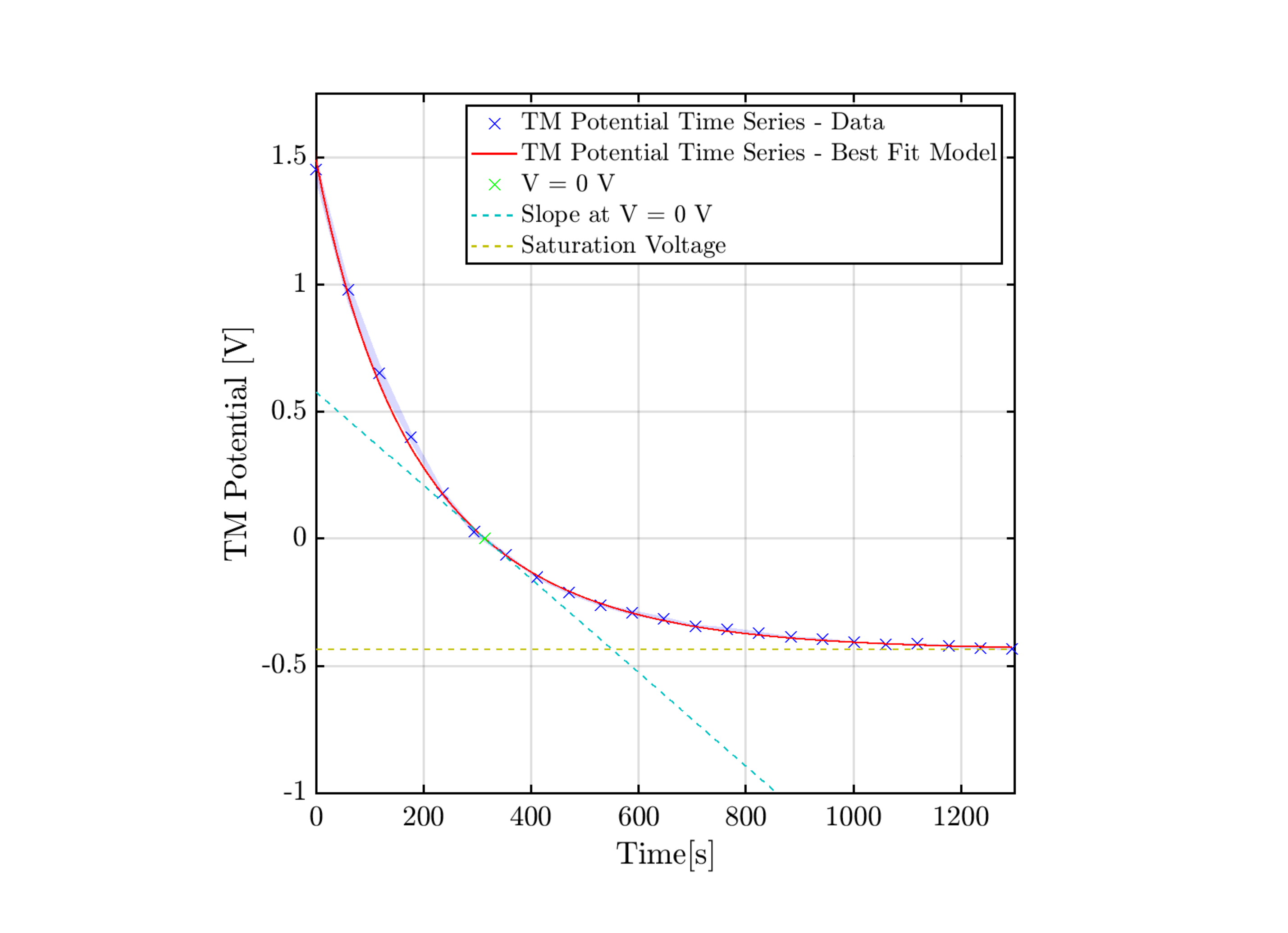}}
    \subfloat[][TM port, $\phi = 36 \si{\degree}, V_{\rm{inj}}=2.05$\,V]{\includegraphics[scale=0.45, trim={4.5cm 1.0cm 4.5cm 0.0cm}, clip]{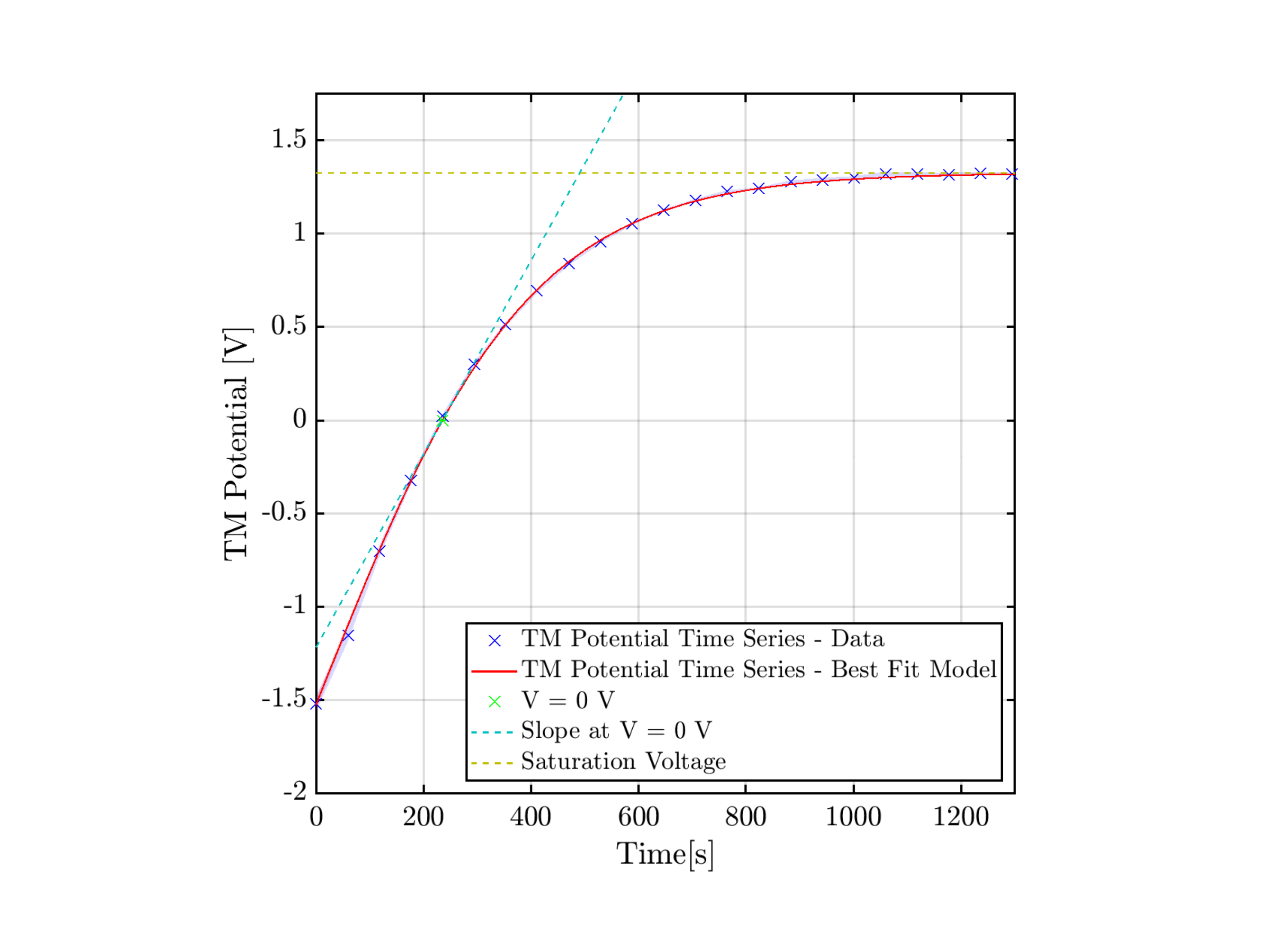}}
    \caption{Time series of the evolution of the TM potential measured during charge measurements on the UF torsion pendulum, as the simplified GRS was illuminated by UV light towards the electrode housing using the EH port (left) and towards the test mass using the TM port (right). The time series data (blue cross points) were fit with the exponential functions given by Equation (\ref{eq: RiccatiEquationSolutions - Potential}) (red solid line), from which were extracted the potential rate at $V_{\rm{inj}} = 0$\,V  (dashed cyan line) and the equilibrium voltage (dashed yellow line).}
\label{figure: TmPotentialTimeSeries}
\end{figure*}


The measurement campaign performed with the torsion pendulum and simplified GRS produced a collection of TM potential time series with one time series for each UV pulse phase setup. In order to extract physical quantities of interest for the two study cases mentioned in the previous section, pre-processing of the time-series was performed. According to the model developed in section \ref{section: Model}, these time-series can be approximated as solutions of ordinary differential equations with second order polynomial source term (also known as Ricatti equations).

\begin{equation}
\frac{\mathrm{d} V_{\rm{TM}}}{\mathrm{d}t} = A V^2 + B V + C,
\label{eq: RiccatiEquation}
\end{equation}
which has the general solution:
\begin{equation}
V_{\rm{TM}}(t) = \frac{d + a\exp(-bt)}{1 + c\exp(-bt)}
\label{eq: RiccatiEquationSolutions - Potential}
\end{equation}
and whose time derivative can be written as:
\begin{equation}
\frac{\mathrm{d} V_{\rm{TM}}}{\mathrm{d}t}(t) = \frac{(bcd - ab)\exp(-bt)}{\big( 1 + c\exp(-bt) \big)^2},
\label{eq: RiccatiEquationSolutions - Potential Rate}
\end{equation}
depending on four parameters, from which one can recognize the equlibrium voltage $V_{\rm{TM}}^{\rm{eq}} = d$ and the time $t^{*} = \tfrac{\ln{(-\sfrac{a}{d})}}{b}$ when the TM voltage crosses 0\,V. The charging rate evaluated at $t = t^{*}$, $\frac{\mathrm{d} V_{\rm{TM}}}{\mathrm{d}t}\Big|_{t = t^{*}}$ then provides the rate when the TM voltage crosses 0\,V.

The measurement campaigns provided such TM potential time series for LED pulse phases, ranging from $0 \si{\degree}$ (injection in-phase, $V_{\text{inj}} = 3.5 V$) to $180 \si{\degree}$ (injection out-of-phase, $V_{\text{inj}} = -3.5 V$). Most were fit with the expression in Equation (\ref{eq: RiccatiEquationSolutions - Potential}). However, for a few phases studied, the TM potential was far from reaching equilibrium at the end of the measurement sequence. 
For such cases, the time derivative of the potential has been assumed linear in $V$ ($A = 0$ in equation \ref{eq: RiccatiEquation}). The general solution therefore becomes a simple exponential function (Equation (\ref{eq: RiccatiEquationSolutions - Potential}) with $c = 0$).

These fits were made using Equation (\ref{eq: RiccatiEquationSolutions - Potential Rate}) and a $\chi^{2}$-minimization with the help of the MATLAB implementation of a Nelder-Mead simplex method \cite{Lagarias1998}. The TM potential rate at 0\,V and the saturation voltage were extracted as functions of the fitting parameters $a$, $b$, $c$ and $d$. The saturation voltage is simply given by the parameter $d$, and the  uncertainty can be derived directly from the fit. The charge rate at 0\,V  
\begin{align}
\frac{\mathrm{d} V_{\rm{TM}}}{\mathrm{d}t}\bigg|_{0\,\rm{V}} & = \frac{(bcd - ab)\exp(-bt^{*})}{\big( 1 + c\exp(-bt^{*}) \big)^2} \nonumber \\
& = \frac{d}{a} \frac{(ab - bcd)}{\big( 1 - \tfrac{cd}{a} \big)^2}
\label{eq: PotentialRateAtZero}
\end{align}
and the uncertainties in these quantities can also be propagated from the fit uncertainties of the parameters $a$, $b$, $c$ and $d$. 
Figure \ref{figure: TmPotentialTimeSeries} illustrates this procedure. It shows: the time evolution of the TM potential as measured with the torsion pendulum (in blue) for the injection phase $\phi = -54 \si{\degree}$, the fit with the exponential model according to Equation (\ref{eq: RiccatiEquationSolutions - Potential}) (in red), the estimated time $t^{*}$ when $V_\indice{TM}$ crosses 0\,V (cross marker in green), the tangent at $t^{*}$ (dashed cyan) and the equilibrium voltage (dashed yellow). The data points in blue come from  a demodulation of the sinusoidal torque excited by the measurement signal.
In the heterodyne demodulation method used, the quadrature component, which contains no signal, provides an estimate of the uncertainty of the demodulation. The error-region about the blue data points in Figure \ref{figure: TmPotentialTimeSeries} are computed from a sliding variance (5-sample window) of the quadrature signal, in order to account for short-term noise variations. The standard deviation is then interpolated across the whole time range to ensure its smooth evolution along the data-set. The errors estimated in this way have been observed to be consistent with the spread of the data points. The typical statistical charge measurement accuracy is around 10\,mV, consistent with the torque noise of the pendulum facility.

An additional, significant source of error was the UV light injection phase. The phase offset between the high frequency sensing voltage and the UV LED pulse was set manually using an oscilloscope. Uncertainty in the UV signal phase arises from limitations of the prototype electronics used to drive the UV LEDs. Future upgrades will include the ability to set the phase of the light relative to the 100\,kHz reference with precise digital timing within an FPGA. For the data reported here, we have estimated that the operator was able to set the injection phase within a $\Delta\phi^{[\si{\degree}]} = \pm 5 \si{\degree}$ accuracy, which corresponds to $\pm 3.5 \si{\volt}\Delta\phi^{[\si{\radian}]} \sin{\big( \phi^{[\si{\radian}]} \big)}$ in terms of electrode voltage. This explains the variation visible in the $x$-axis error bars in Figure \ref{figure: dVdt_Total_PortEh}.
With UV light completely in-phase or out-of-phase with the voltage, ($\phi^{[\si{\degree}]} = 0\si{\degree}$, $\phi^{[\si{\degree}]} = -180\si{\degree}$) the resulting error in voltage is negligible, at $\phi^{[\si{\degree}]} = 90\si{\degree}$, the error is maximal.

\subsection{Electron flow rate as a function of surface potentials}
\label{subsection: Electron flow rate as a function of surface potentials}

\begin{figure*}
    \centering
    \subfloat[][Observed S-curves described by the data-points in blue (extracted from the potential time series fit in Figure \ref{figure: TmPotentialTimeSeries} for each phase) and the model adjustment in red.\label{figure: dVdt_Total_PortEh}]{\includegraphics[scale=0.45, trim={0.0cm 0.5cm 0.0cm 0.0cm}, clip]{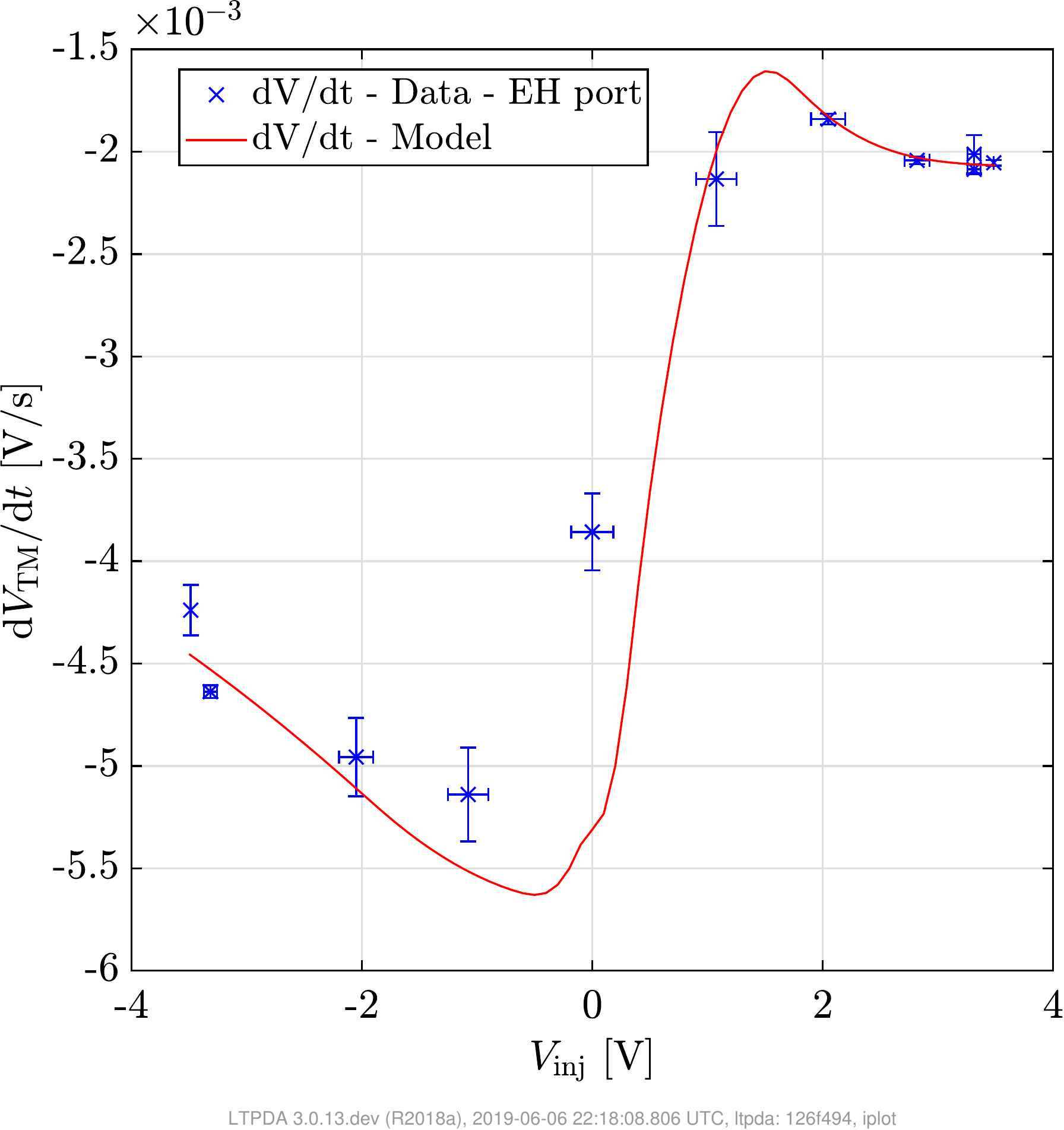}}
    \hspace{10mm}
    \subfloat[][Decomposition of the net flow over the contributions from the four flows of electrons existing between the four relevant surfaces.\label{figure: dVdt_ModDecomp_PortEh}]{\includegraphics[scale=0.45, trim={0.0cm 1.0cm 0.0cm 0.0cm}, clip]{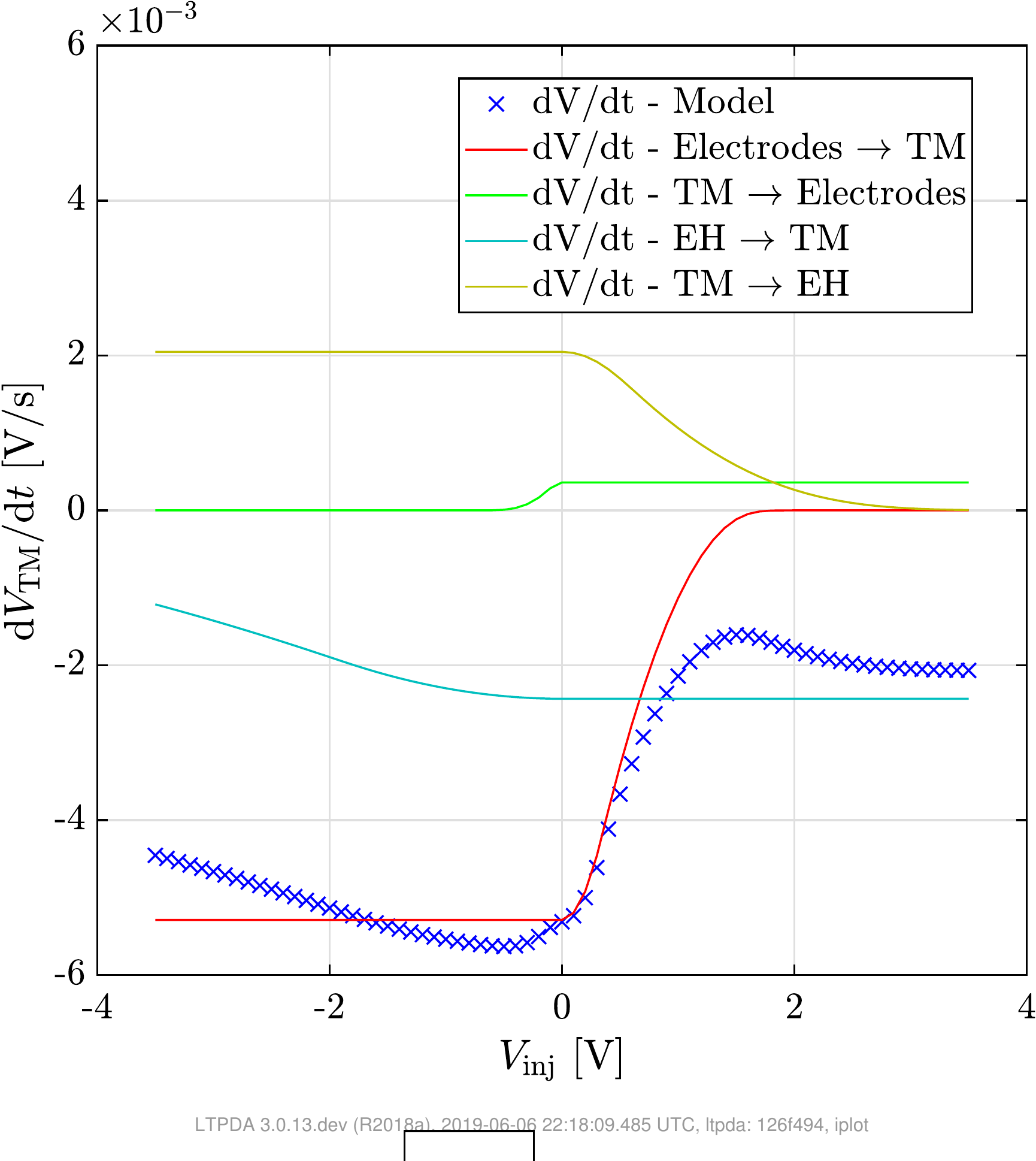}}\\
    \subfloat[][Observed S-curves described by the data-points in blue (extracted from the potential time series fit in Figure \ref{figure: TmPotentialTimeSeries} for each phase) and the model adjustment in red.\label{figure: dVdt_Total_PortTm}]{\includegraphics[scale=0.45, trim={0.0cm 0.5cm 0.0cm 0.0cm}, clip]{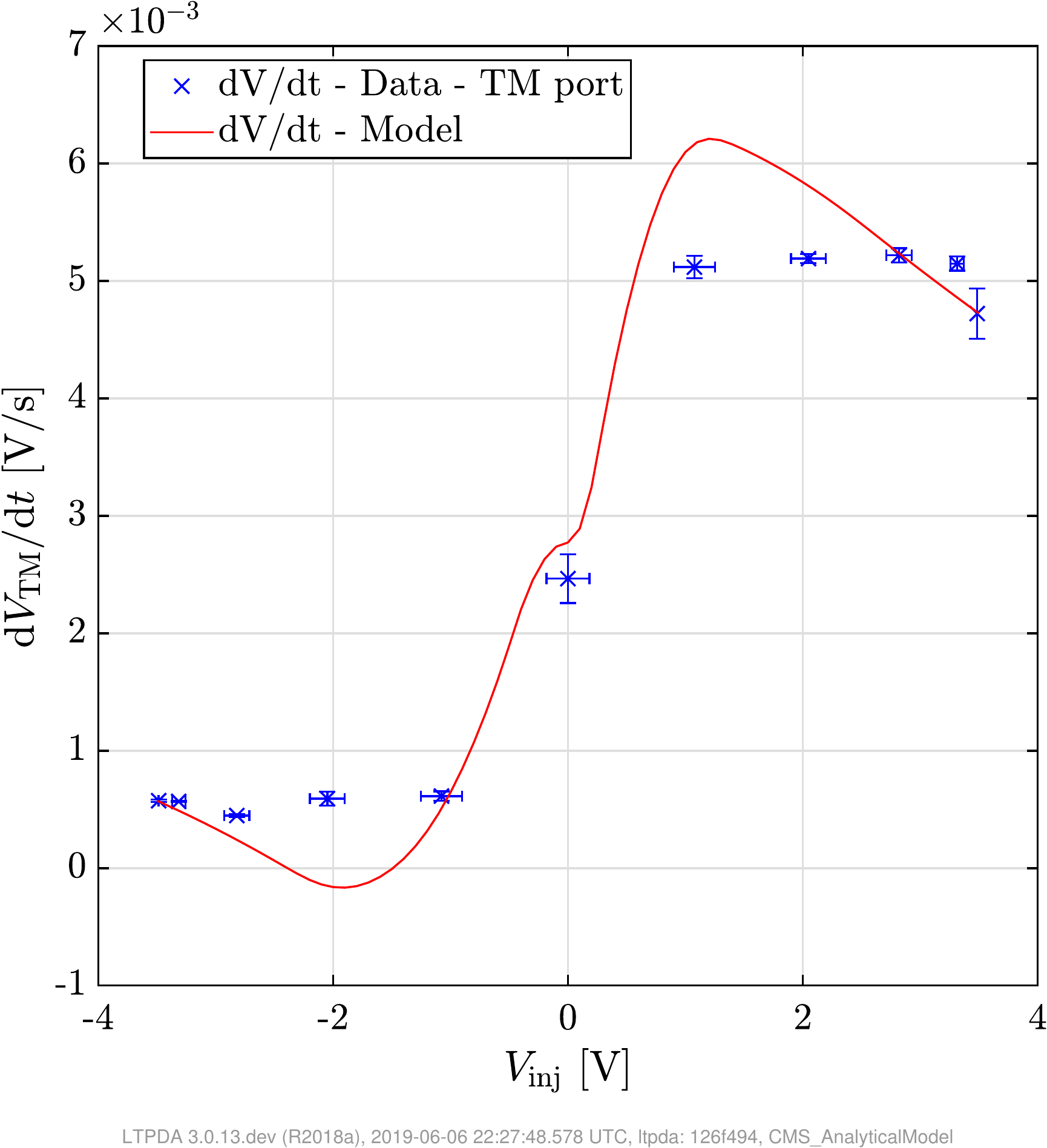}}
    \hspace{10mm}
    \subfloat[][Decomposition of the net flow over the contributions from the four flows of electrons existing between the four relevant surfaces.\label{figure: dVdt_ModDecomp_PortTm}]{\includegraphics[scale=0.45, trim={0.0cm 1.0cm 0.0cm 0.0cm}, clip]{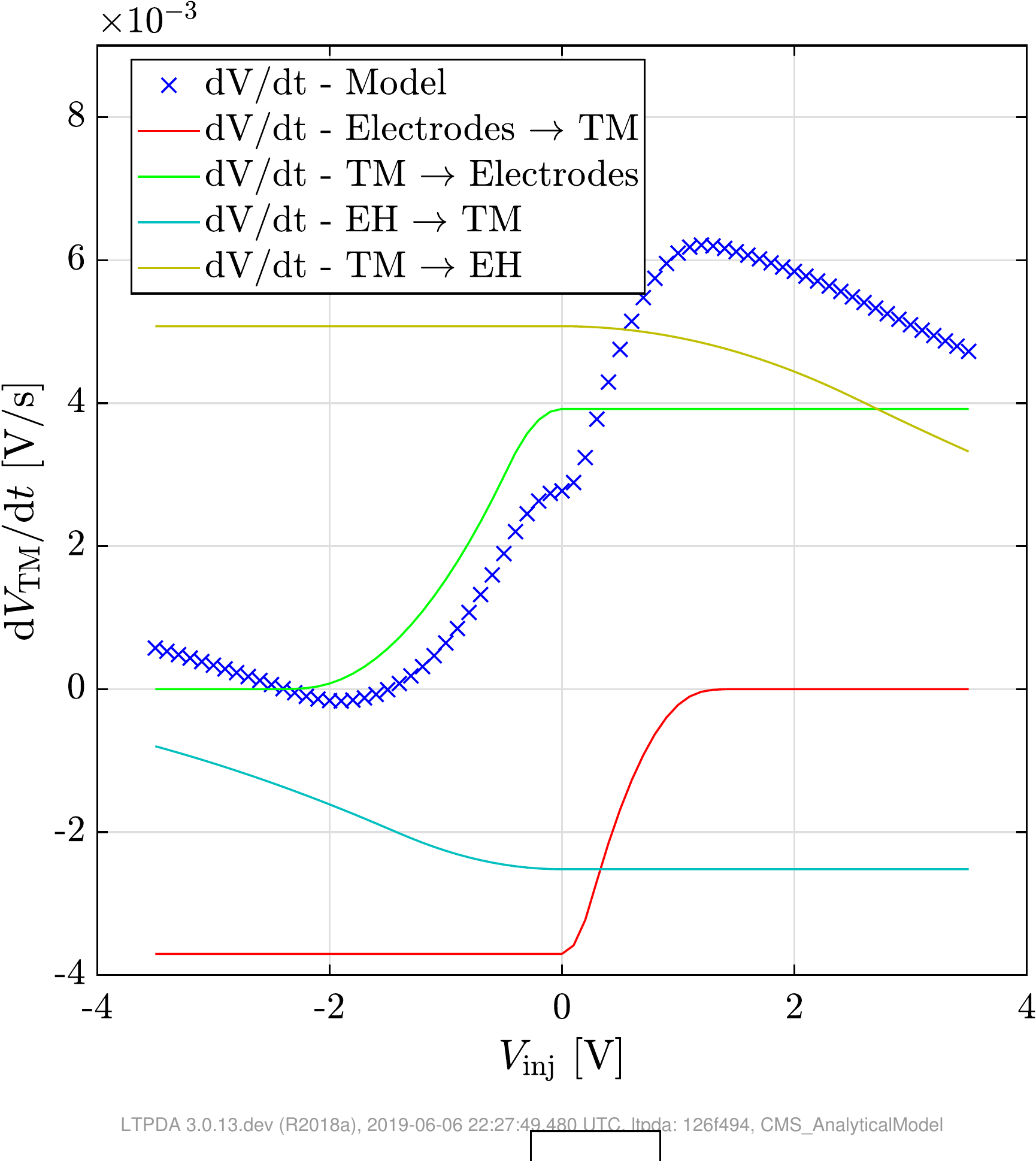}}
    \caption{TM potential rate at $V_{\rm{TM}} = 0 \si{\volt}$ as a function of the applied voltage $V_{\rm{inj}}$, in the case of a EH illumination and TM illumination.}    
    \label{figure: dVdt_Total}
\end{figure*}

The model predictions presented below originate from the joint fit of the parametric model described in section \ref{section: Model} using two sets of data: the TM charging rates (Figures \ref{figure: dVdt_Total}\subref{figure: dVdt_Total_PortEh} and \ref{figure: dVdt_Total}\subref{figure: dVdt_Total_PortTm}) and the equilibrium voltages in Figure \ref{figure: Vsat}, in a joint analysis. Hence, both data sets have to be considered in order to judge the goodness of the fit. The fit itself is the result of a $\chi^{2}$ pre-minimization with the Nelder-Mead simplex method \cite{Lagarias1998} implemented in MATLAB with the fmincon() function, which allows us to constrain parameters to a certain interval during the minimization process. For instance, work function parameters are restricted to between 3.4 and 4.8\,eV following previous experimental results involving gold-coated test masses \cite{Armano2018a, Wass2019}. This pre-optimization is used as the initialization step of a Markov-Chain Monte Carlo (MCMC) sampling \cite{Foreman-Mackey2013} of the posterior density function. The MCMC allows for a comprehensive study of the parameter space, and for a robust extraction of the fit uncertainties using the non-linear model parameters. The MCMC optimization provides the parameters used to evaluate the fit models presented below.

The model has been evaluated with the fit results for the evolution of the TM charge rate at $V_{\rm{TM}} = 0$\,V as the applied potential $V_{\rm{inj}}$ is varied (via the phase between the LED pulses and the 100\,kHz injection signal).
Figure \ref{figure: dVdt_Total}\subref{figure: dVdt_Total_PortEh} presents the measurements and the model predictions in the case where the EH port is used for illumination. The measured data points are presented (blue dots), together with the model predictions (solid red line), while Figure \ref{figure: dVdt_Total}\subref{figure: dVdt_ModDecomp_PortEh} presents the decomposition of the total TM potential rate into the four current flows within the two contributing capacitors, according to Equation (\ref{eq: ndot}).

To build the $\chi^{2}$ function including $x$-axis error-bars, projection onto the $y$-axis was necessary, leading to an increased complexity of the $\chi^{2}$ topology. Indeed, the operation $\Delta y = \pm g(\Delta x)$ is model-dependent. An iterative strategy was adopted to address this issue. The first step ignored the $x$-axis error-bars and yielded an initial parameter set from which the model fit was computed providing values and slopes at each $x$ value. This resulting model was then used to convert injection voltage uncertainty to error in the potential rate according to Equation \ref{eq: DxToDy} below, and another set of data with corrected $y$-axis error-bars was produced. The procedure was repeated $10$ times, sufficient to observe convergence of the algorithm
\begin{equation}
\Delta y = \pm g(\Delta x) = \pm \bigg| \frac{f(x + \Delta x) - f(x - \Delta x)}{2} \bigg|.
\label{eq: DxToDy}
\end{equation}

The model predictions traced in red in Figure \ref{figure: dVdt_Total_PortEh} (left), and its decomposition (right) provide a physical understanding of the S-shaped curves observed in the measurements of $\frac{\mathrm{d}V_{\rm{TM}}}{\mathrm{d}t}$ against $V_{\rm{inj}}$. 

Each contribution exhibits a polynomial shape from $\Delta V=0$ to $V_m$ according to Equations (\ref{eq: ElEnDist}) and (\ref{eq: emissionRatio}).
The work function and distribution of electron energies emitted from each surface define the energy range over which the rising or falling portion of the curves extend and their slope. 
Unfortunately the number of measurements points constraining these parameters is limited to only a few points near $V_{\rm{inj}}=0$\,V due to the combination of the difficulty in setting the phase accurately close to the zero-crossing of the 100\,kHz injection signal and the fact that $h\nu - \Phi$ is small compared to the maximum $V_{\rm{inj}}$.
Future experimental campaigns can improve on this by more accurately setting the phase of the UV illumination. 

In the central region around $V_{\rm{inj}}=0$, there is a steep change in charge rate as a function of $V_{\rm{inj}}$, with a width defined as expected by $h\nu-\Phi$ produced by the electron flow between test mass and injection electrode.
On the left and right end of the charge-rate curves, shallower slopes are  observed, driven by the current flow between TM and EH surfaces. Here the rate of change of $\mathrm{d}V_{\rm{TM}}/\mathrm{d}t$ with respect to $V_{\rm{inj}}$ is lower because the potential difference between the surfaces is produced by the capcacitively induced test mass potential at 100\,kHz $\Delta V_{\rm{tm-eh}} = \eta V_{\text{inj}}$. 

The piece-wise nature of the photo-electron energy distribution used in our model, introduces discontinuities in the slope of the model prediction of charging rate, especially close to $V_{\rm{inj}}=0$\,V. Although the fit to the overall shape of the curve is good, there are insufficient data to constrain the shape of the energy distribution.

Figures \ref{figure: dVdt_Total}\subref{figure: dVdt_Total_PortTm} and \ref{figure: dVdt_Total}\subref{figure: dVdt_ModDecomp_PortTm} show the measured and modelled curves for UV illumination directed at the TM. The TM charge rate is observed to be positive across the whole curve, showing a net transfer of electrons from TM to EH.
The physical breakdown of the net flow exhibits similar features to the EH illumination: a steep slope in a narrow voltage interval around 0\,V driven by electrons flow between the 100\,kHz electrode and the TM surface, and shallower slope produced by the induced voltage on the TM relative to the grounded EH. However, for this UV port, there is a discrepancy between the best fit model in red, and the data points in blue. At the extremes of the curve, the data favors a constant charge rate as function of $V_{\rm{inj}}$.  
The physical interpretation of this discrepancy is difficult to identify -- a lower value of work function, $\Phi$, would result in a weaker voltage dependence but would need to be local to the illuminated areas of the test mass and electrode housing outside of the 100\,kHz electrode. An electron energy distribution with a peak weighted at higher energy would also lower the slope of the curve. Again this effect would need to be present only in the surfaces illuminated between TM and EH. Interestingly, some of these surfaces are located around the corners and edges of the test mass where the geometry of emission favors such a change. Finally, stray potential offsets between surfaces can shift the electron energy distribution to higher energies, extending the flat portion of the charge rate contribution curves. 
Despite this slight disagreement between measured charge rate data and the model fit for the TM illumination, equilibrium voltage measurements for the same illumination shown in  on the right-hand plot of Figure \ref{figure: Vsat}, produced by the same model and which will be discussed further in the next section show good agreement. 

\subsection{Equilibrium potential as a function of applied voltage}
\label{subsection: Saturation voltage as a function of applied voltage}

 Several proposed modes of operation of the CMS rely on the ability to adjust the test mass equilibrium potential of the TM close to 0\,V.
 A better understanding of the physics of this equilibrium will allow for appropriate tuning of the parameters under control, most importantly the optimal value of $V_{\text{inj}}$ and the preferred ports of illumination. Our model provides a first modeling of those dependencies.
 
\begin{figure*}
    \centering
    \subfloat[][TM illumination case\label{figure: Vsat_Tm}]{\includegraphics[scale=0.45, trim={0.0cm 0.5cm 0.0cm 0.0cm}, clip]{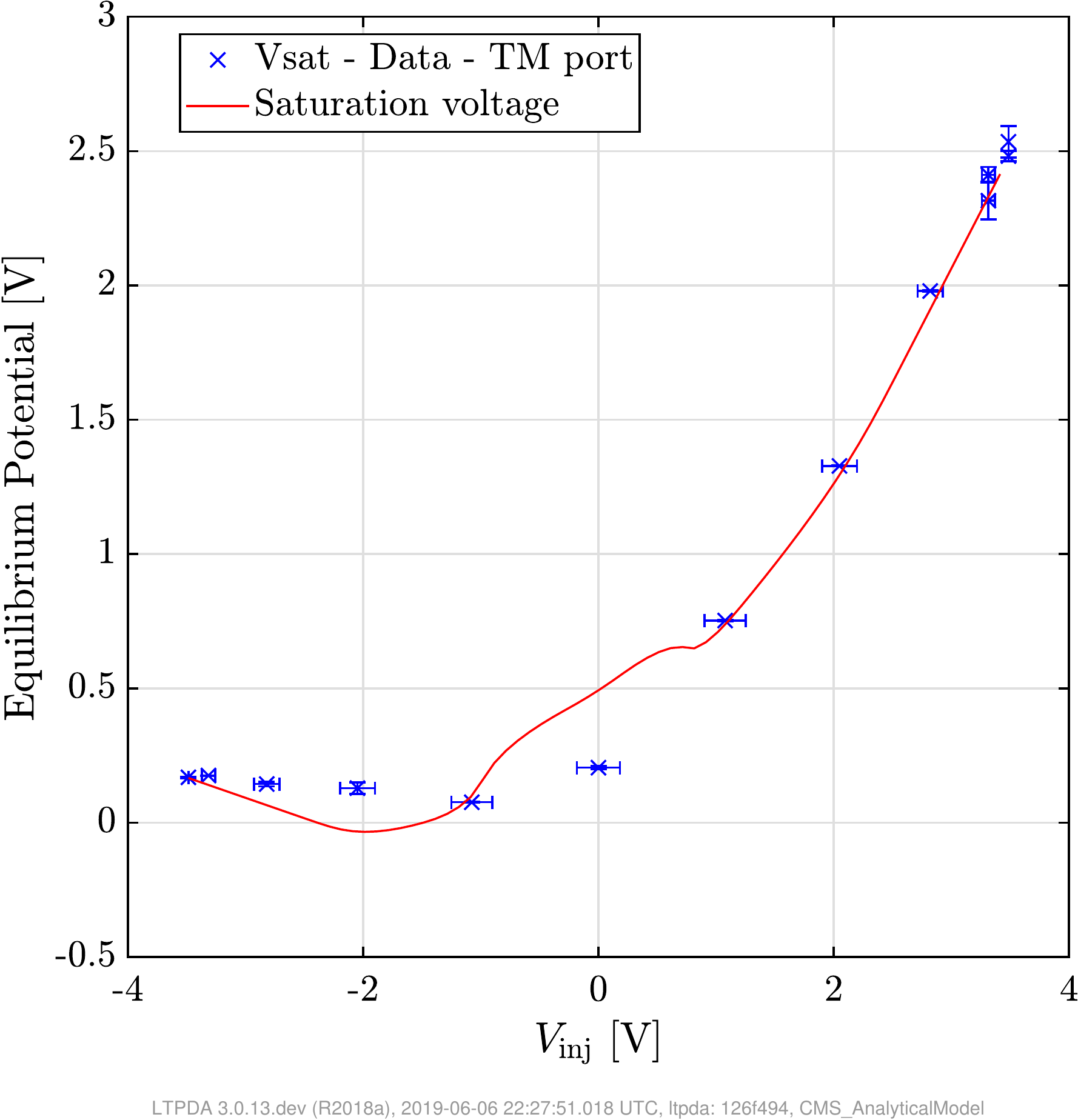}}
     \hspace{10mm}
      \subfloat[][EH illumination case\label{figure: Vsat_Eh}]{\includegraphics[scale=0.45, trim={0.0cm 0.5cm 0.0cm 0.0cm}, clip]{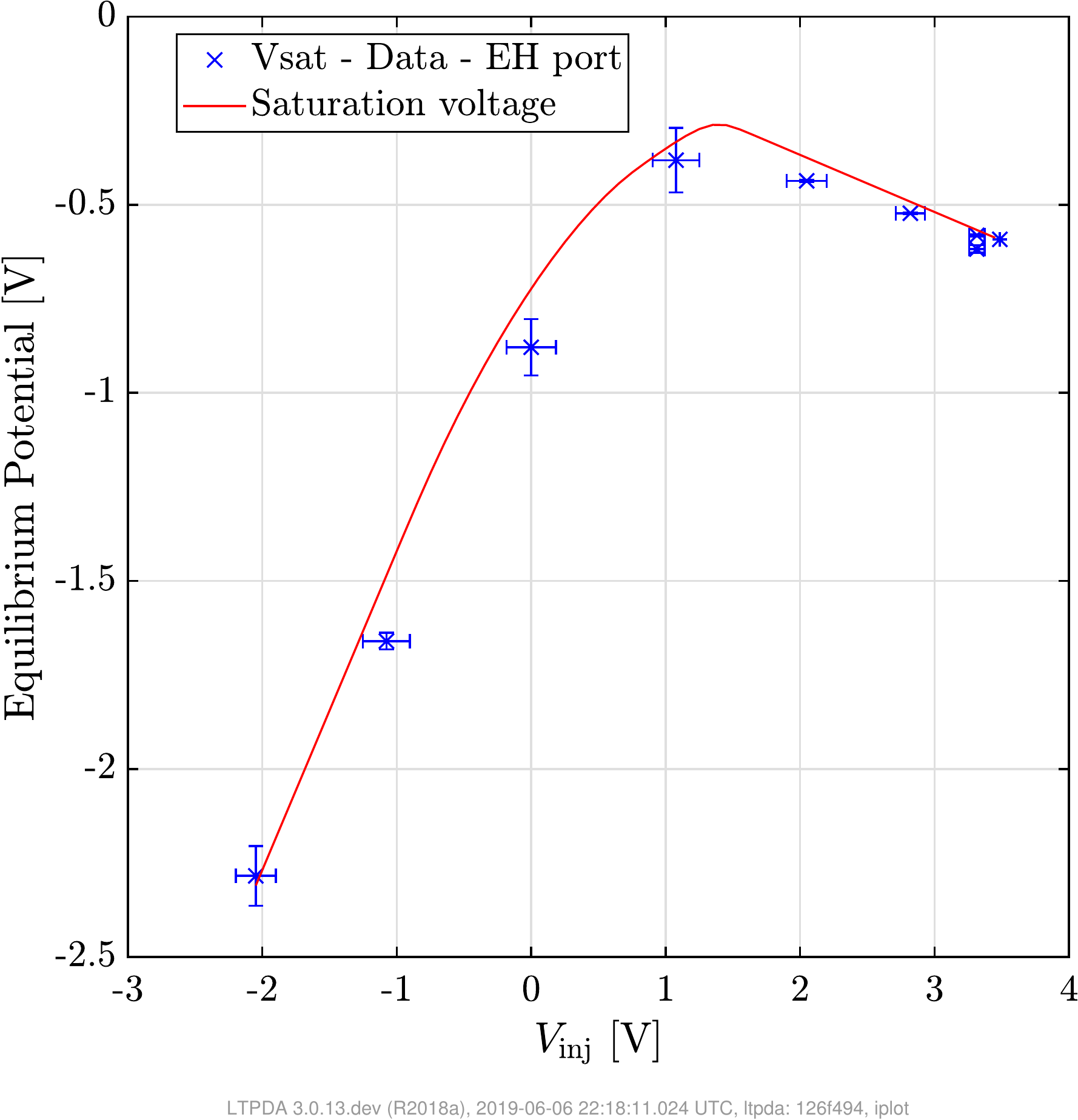}}
    \caption{Measured saturation voltage versus the applied voltage $V_{\rm{inj}}$. The blue data-points represent the measurements (extracted from the potential time series fit e.g. in Figure \ref{figure: TmPotentialTimeSeries} for each phase) and the red curves show the best-fit model.}
\label{figure: Vsat}
\end{figure*}

 Measurements of the TM equilibrium potential as a function of $V_{\rm{inj}}$ provide complementary information to measurements of the charge rate at $0\ V $, providing independent constraints on the model parameters and potentially removing degeneracies. Figures \ref{figure: Vsat}\subref{figure: Vsat_Eh} and \subref{figure: Vsat_Tm} show the saturation voltage measured for various injection voltage $V_{\text{inj}}$ value for both EH illumination and TM illumination. Figure \ref{figure: Vsat} also shows the results of the combined fit, showing the model predictions compared with the measured equilibrium voltages. In general the fit to the data is good and follows the behavior described in detail in Section \ref{section: Model}. The results for the EH port are similar to illumination case 1 in Figure \ref{fig: EqPotential_SimpleModel}, with $V_{\rm{eq}}$ varying in proportion to $V_{\rm{inj}}$ for negative $V_{\rm{inj}}$ and transitioning to a $-\eta V_{\rm{inj}}$ dependence for positive $V_{\rm{inj}}$. The TM port shows the reverse behavior, varying as $-\eta V_{\rm{inj}}$ for negative $V_{\rm{inj}}$ and proportional to $V_{\rm{inj}}$ for positive similar to case 2 in Figure \ref{fig: EqPotential_SimpleModel}. 


\subsection{Measuring model parameters and tracking photons}
\label{subsection: Measuring model parameters and tracking photons}

From the Equations (\ref{eq: Net potential rate}) and (\ref{eq: ndot}) describing the simple analytical model, it is clear that the physical scaling parameters: the UV power input to the system, $P_{\rm{UV}}$, the illumination fractions $\alpha_{i}$ or the Quantum Yield are inseparable in the model fit. 
Only the four combinations $P_{\rm{UV}} QY \alpha_{i}$ and the two work functions of the TM and EH surface materials can be extracted. 
However, a quantitative measure of the $\vec{\alpha}$ can be obtained using external estimates of $P_{\rm{UV}}$ and $QY$. 

The UV power at the input of the vacuum chamber port has been measured for all measurements shown here, combined with the fiber optic transmission inside the vacuum chamber, the power reaching the sensor was $P_{\rm{UV}} = 150 \pm 10 \si{\nano\watt}$ for each measurement run. The quantum yield of similar gold surfaces has been studied in \cite{Hechenblaikner2012, Armano2018a, Sumner2020, Olatunde2018}. In systems which are allowed to stabilize over long periods (~weeks-months), the quantum yield of has been found in several circumstances to approach $\sim2\pm0.5\timestentothe{-5}$ for a light source with a similar wavelength as used here. 

Applying these values, we determine the illumination fractions when using the electrode housing port, $\vec{\alpha}_\indice{\text{EH}}$, and when using the test mass port, $\vec{\alpha}_\indice{\text{EH}}$:
\begin{align}
\vec{\alpha}_\indice{\text{EH}} = 
\begin{bmatrix}
0.37 \pm 0.03 \\
0.03 \pm 0.05 \\
0.17 \pm 0.05 \\
0.14 \pm 0.002
\end{bmatrix}
&&
\vec{\alpha}_\indice{\text{TM}} = 
\begin{bmatrix}
0.26 \pm 0.03 \\
0.28 \pm 0.02 \\
0.18, \pm 0.01 \\
0.36 \pm 0.02
\end{bmatrix}
\label{eq: MeasuredAlpha}
\end{align}
and the work functions when using the electrode housing port, $\vec{\Phi}_\indice{\text{EH}}$, and when using the test mass port, $\vec{\Phi}_\indice{\text{TM}}$:
\begin{align}
\vec{\Phi}_\indice{\text{EH}} = 
\begin{bmatrix}
3.6 \pm 0.19 \\
4.6 \pm 0.05
\end{bmatrix}
&&
\vec{\Phi}_\indice{\text{TM}} = 
\begin{bmatrix}
3.96, \pm 0.16 \\
3.10 \pm 0.09
\end{bmatrix}
\label{eq: MeasuredWorkFunc}
\end{align}

The illumination estimates can be verified to some degree by a simple ray-tracing estimate. While a full light-tracking simulation such as those described by \cite{Armano2018a, Ziegler2014} are beyond the scope of this paper, simple geometric arguments can provide estimates that can be compared with the fit results.
To this end we make use of a cross section through the CAD model of the simplified GRS in the $y$-$z$ plane at the $x$ position of the fiber injector as shown in Figure \ref{fig:UV LED geometries}. The following assumptions are made about the UV ray propagation: all reflections are specular, the angular distribution of the light output from the fiber injector is a Gaussian with a 2-$\sigma$ half-width of 12.5-degrees consistent with the numerical aperture of the multimode fiber of 0.22, the reflectively of gold is taken from Reference \cite{Johnson1972}. We restrict our analysis to the 2-D plane described above and take into account the first four reflections only.  

\begin{figure}
    \centering
    \centerline{\includegraphics[scale=0.28, trim={0.5cm 0.3cm 0.3cm 0.1cm}, clip, frame]{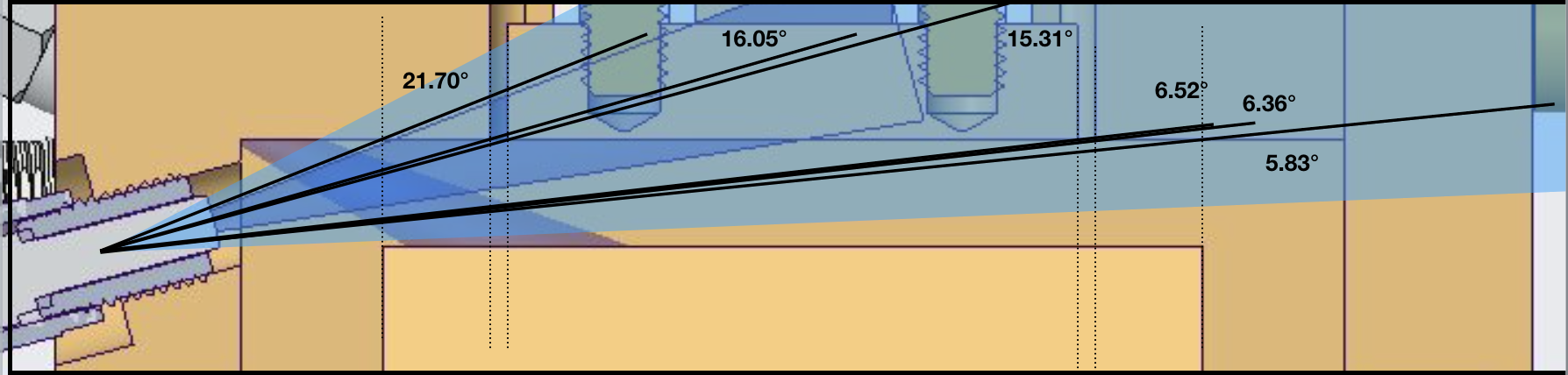}}
    \vspace{0.2cm}
    \centerline{\includegraphics[scale=0.28, trim={0.5cm 0.3cm 0.3cm 0.1cm}, clip, frame]{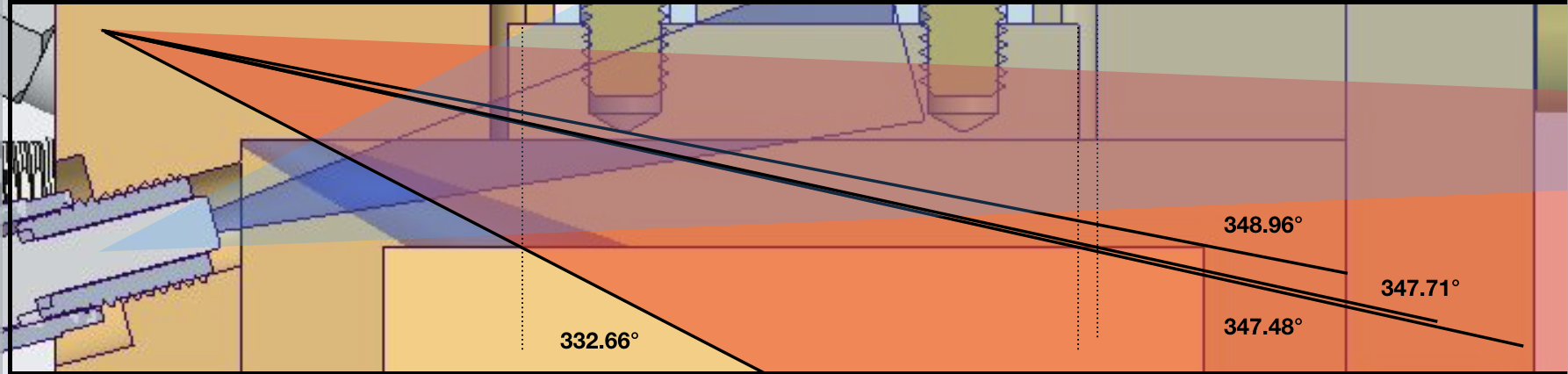}}
    \vspace{0.2cm}
    \centerline{\includegraphics[scale=0.28, trim={0.5cm 0.3cm 0.3cm 0.1cm}, clip, frame]{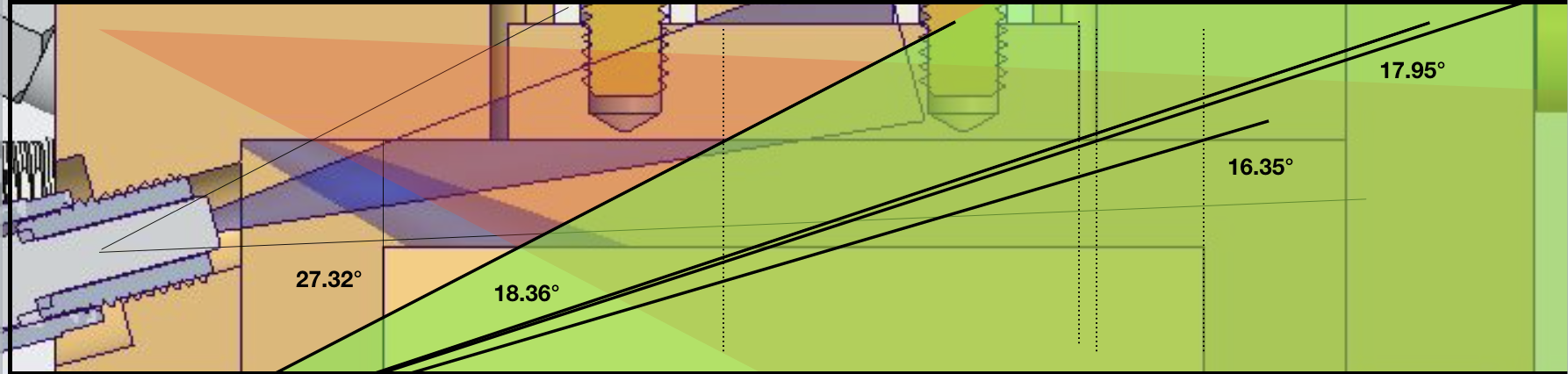}}  
    \vspace{0.2cm}
    \centerline{\includegraphics[scale=0.28, trim={0.5cm 0.3cm 0.3cm 0.1cm}, clip, frame]{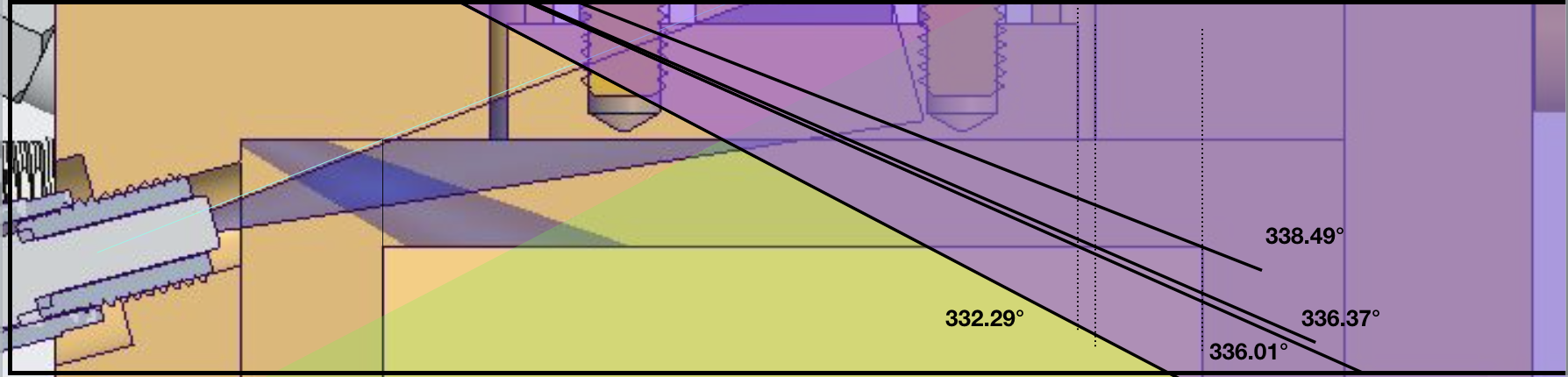}}
    \caption{A simple photon tracking study using the CAD model of the simplified GRS. Here illumination through the EH port is considered. For each five successive cones (blue, red, green, purple and white) from the incident and the reflected light, the UV beam is divided according to the sections of the light cone received by each category of surfaces (TM opposite the EH, TM opposite the electrode, electrode, and EH). The angles defining these fractions are measured and fed into the calculation, Equation (\ref{eq: ReflectionStudy}), that estimates the parameters $\vec{\alpha}$}
    \label{figure: LightReflections}
\end{figure}

We identify the four categories of surface of interest inside the GRS (TM opposite the EH, EH, test mass opposite the electrode, and electrode). The light cone is projected from the UV port of interest and a measure of the angles subtended by the edges of each surface is made (the measured angles are shown in Figure \ref{figure: LightReflections}, for the successive reflections considered). The light intensity incident on each surface is calculated integrating the angular intensity distribution between the measured limits. 

The UV light is propagated to a second reflection, defined by the angle of incidence of the central and extremal rays of the beam. The reflectivity is determined by the Fresnel equations and the complex index of refraction of gold. The reflected rays define the clipped Gaussian beam impinging on the opposing surfaces and the intensity on the opposing surfaces of each type is calculated in the same way as the primary beam. 

These considerations allow one to build an expression for the quantity of UV light absorbed with reflection number $i$ on a given surface as:
\begin{align}
& \alpha_{i} = N_{i}^{\text{inc}} \Bigg[ 1 - \frac{\int_{\theta_{1}}^{\theta_{2}} g(\theta) R(\theta)^{(i)} d\theta}{\int_{\theta_{1}}^{\theta_{2}} g(\theta) R(\theta)^{(i - 1)} d\theta} \Bigg] \\
& N_{i}^{\text{inc}} = N_{0}^{\text{inc}} - \sum_{1}^{i - 1} \alpha_{i} \\
& N_{0}^{\text{inc}} = \int_{\theta_{1}}^{\theta_{2}} g(\theta) d\theta
\label{eq: ReflectionStudy}
\end{align}
where $g(\theta)$ is the Gaussian profile of the light beam, $R(\theta)$ is the reflectivity as a function of angle, $\theta_{1}$ and $\theta_{2}$ are the incident angles of the two extreme rays of the part of the beam incident on the surface to be calculated and $i$ the index of the current bounce.

From Equation (\ref{eq: ReflectionStudy}) and the measured angles from Figure \ref{figure: LightReflections}, we obtain the following values for $\vec{\alpha}$. 
\begin{align}
\vec{\alpha}_\indice{\text{EH}} &= \big[ 0.36, 0.02, 0.22, 0.06 \big] \nonumber \\
\vec{\alpha}_\indice{\text{TM}} &= \big[ 0.30, 0.26, 0.02, 0.27 \nonumber \big]
\end{align}

Comparing with the parameters obtained by the model fit to the measured data, this photon tracking analysis shows good agreement. The EH port case yields the best result, agreeing well across all four surface categories. The $\vec{\alpha}_\indice{\rm{EH}}(4)$ value, referring to the light absorbed by tm-eh, shows the most significant discrepancy between the value obtained by ray tracing and from the fit, the measured value being twice as big as the estimate.
Uncertainties in the reflection modelling are largest for the surfaces illuminated farthest from the UV port.
The reflection modeling stops after considering $5$ reflections, and does not take into account propagation of light reflected along the perpendicular faces of the test mass. Constraining the analysis to a single plane becomes an increasingly poor approximation as the the light traverses the test mass and it spreads into the $x$-$y$ plane.
The tm-eh surfaces reached by the light from the EH port are indeed near the opposite housing corner, far from the injection port. 
In the TM port case, the parameters $\vec{\alpha}_\indice{TM}(3)$ and $\vec{\alpha}_{\rm{TM}}(4)$ show a discrepancy between the two estimation methods. The TM port illumination estimated from the fit yields significantly more light on the eh-tm surface ($\vec{\alpha}_{\rm{TM}}(3)$ parameter) than the ray-tracing. The ray-tracing predicts that the light reaches this surface after the third reflection only, at which point the uncertainty in the ray-tracing calculation has grown large. 

The work function values returned by the fit are given in Equation (\ref{eq: MeasuredWorkFunc}). While the work function value for the EH surfaces measured with EH and TM ports illumination are consistent at the $1 \sigma$ level, the values for TM surfaces are in tension. 
It is possible that different regions of the sensor surfaces have different surface properties, as has been suggested by previous work \cite{Armano2018a}. However, we suspect the fit to the TM illumination charge measurements may be biased by the difficulty in fitting the data described in Sections \ref{subsection: Electron flow rate as a function of surface potentials} and \ref{subsection: Saturation voltage as a function of applied voltage}. In particular, the fit may have systematically favored a low value for the TM work function in order to explain the observed low dependence of the charge rate on $V_{\rm{inj}}$ for larger positive and negative values of $V_{\rm{inj}}$ in Figure \ref{figure: dVdt_Total_PortTm}.

\section{Conclusion}

We have presented the first measurements of contact-free UV discharging by photoemission synchronized with AC electric fields in a gravitational reference sensor relevant for ultra-sensitive space-based gravitational missions. We make use of a deep UV LED light source which is a candidate technology for the future gravitational wave mission, LISA. Further, we have demonstrated the ability to control the GRS test mass charge rate and potential by adjusting the phase of 100\,kHz UV light pulses with respect to the AC field. By taking advantage of AC electric fields and preferential illumination of the test mass and sensor surfaces, this method allows the charge rate and potential of the test mass to be controlled over a broad range. In a DC illumination scheme as demonstrated during the LISA Pathfinder mission \cite{Armano2018a}, similar levels of charge control required applied voltages on GRS electrodes that introduce unwanted force disturbances. 
Our torsion pendulum facility has allowed us to probe the discharging behavior using representative hardware in a relevant environment. The precision of our measurements is comparable to the state of the art in other terrestrial laboratories and within 2 orders of magnitude of that demonstrated in space with LISA Pathfinder.

We have developed a simple model of the discharge process taking into account the two relevant electric field regions in our simplified GRS. The model is simpler and more computationally efficient than previous work addressing the design of the LISA Pathfinder sensor \cite{Ziegler2014}. Instead the principle of our model is similar to the recent work that addresses LISA Pathfinder in-flight discharging results \cite{Armano2018a}: both  reduce the GRS to a system of parallel plate field regions. We extend this treatment to consider UV light injection synchronized with instantaneous fields. Our sensor is simpler in its electrode geometry than the LISA Pathfinder flight model but the model is easily extended to include a larger number of electrodes. 
Although conceptually simple, our model successfully describes the test mass equilibrium potentials in the system and how they depend on the UV illumination properties and relative phase of the pulsed UV light. Given the surface properties, and light distribution within the sensor, the model explains well our measurements of test mass discharge rate and equilibrium shifts.

We have developed a data analysis and model-fitting framework that allows us to determine the properties of the discharging system based on our physical model. Using a comprehensive set of measurements as inputs, the fitting process combines all available information to produce a single set of parameters for the system. The best fit model to the data, explains the measured behavior of the system. Using reasonable assumptions for the quantum yield properties of the GRS that have not been measured, the results of the model fit describing the UV illumination distribution within the sensor agree well with a simple ray-tracing analysis. 

Both our physical model and data analysis framework are extendable to future work. A key uncertainty in understanding the discharging properties of similar systems is the unpredictability and variability of surface quantum yields and work functions, especially under non-ideal vacuum conditions necessary for space missions (bake out in the presence of volatile contaminants, leaking to rough vacuum during launch preparations) and this has been studied in dedicated tests at sample level \cite{Wass2019, Olatunde2018}. If the illumination distribution can be determined through an independent analysis as in the case of \cite{Ziegler2014, Armano2018a} then the analysis described here provides a robust method for obtaining accurate determination of the sensor surface properties. This can be used to monitor and explain changes measured for example after vacuum bake out with all other conditions equal.

This work is directly applicable to the development of the LISA discharge system. In our torsion pendulum facility we have recently installed a new GRS with a LISA-like geometry including 12 electrodes for six-degree-of-freedom sensing and actuation. We will use this sensor to explore synchronized discharging with UV injection schemes which are designed to  take maximum advantage of the strongest electric fields in the sensor (similar to those described in \cite{Ziegler2014}). We also anticipate the application of LISA discharging technology in GRS for future Earth geodesy missions, using drag-free test masses to enhance the acceleration noise performance of inertial sensors.

\newpage

\section{Appendix}
\label{section: Appendix}

The electron flow rate related to the electron energy distribution, as described in Section \ref{section: Model}, has the general form shown in Equation \ref{eq: A1}. 

\begin{center}
\begin{equation}
\dot{n} = \alpha\ QY\ W \frac{\ P_{\rm{UV}}}{ h \int_{0}^{+\infty}\nu g(\nu) d\nu} \int_{\frac{\phi}{h}}^{+\infty} g(\nu) d\nu 
\label{eq: A1}
\end{equation}
\end{center}
In the case of an infinite parallel plate geometry (blue curve in Figure \ref{fig: electDistSimple}), $W$ is defined as $W_b$ in Equation (\ref{eq: A2}),
while in the case accounting for a finite plate approximation used in our situation (red curve in Figure \ref{fig: electDistSimple}), $W$ is defined as $W_r$ in Equation (\ref{eq: A3}).

\begin{center}
\begin{equation}
\ W_b= \frac{\int_{\frac{e\Delta V + \phi}{h}}^{+\infty} g(\nu) d\nu \int_{\Delta V}^{V_m} f(\phi,\Delta V,\nu,T) d\Delta V}{\int_{\frac{\phi}{h}}^{+\infty} g(\nu) d\nu\int_{0}^{V_m} f(\phi,\Delta V,\nu,T) d\Delta V}
\label{eq: A2}
\end{equation}
\end{center}

\begin{equation}
W_r=\dfrac{\splitdfrac
{(\int_{\frac{e\Delta V + \phi}{h}}^{\frac{e \frac{\Delta V}{x} + \phi}{h}} g(\nu) + \int_{\frac{e\Delta V + \phi}{h}}^{+\infty} g(\nu))d\nu }{(\int_{\Delta V}^{x V_m} f(\phi,\Delta V,\nu,T)+\int_{x V_m}^{V_m} f(\phi,\Delta V,\nu,T))d\Delta V}}{\int_{\frac{\phi}{h}}^{+\infty} g(\nu) d\nu\int_{0}^{V_m} f(\phi,\Delta V,\nu,T) d\Delta V}
\label{eq: A3}
\end{equation}

Given that the photon energy, $g(\nu)$, is well described by a normal distribution with mean $\overline{\nu}$ and standard deviation $\sigma$, resolving the integrals in Equations 21, 22 and 23 can be made easier by defining the following,
\begin{align}
& A = \int_\frac{e\Delta V +\phi}{h}^{+\infty} g(\nu) d\nu
\label{eq: A4}
\end{align}
\begin{align}
& B = \int_\frac{e\Delta V +\phi}{h}^{+\infty} \nu g(\nu) d\nu
\label{eq: A5}
\end{align}
\begin{align}
& C = \int_\frac{e\Delta V +\phi}{h}^{+\infty} {\nu}^2 g(\nu) d\nu
\label{eq: A6}
\end{align}
Making substitutions for $A$, $B$ and $C$ in Equation (\ref{eq: A2}) results in,
\begin{align}
    W_b = \frac{e^2\Delta V A + (2e \phi A - 2ehB)\Delta V + h^{2}C - 2h\phi B + \phi^{2}A}{h^{2}C - 2h\phi B + \phi^{2} A}
    \label{eq: A7}
\end{align}
where,
\begin{align}
   \ eV_{m} = \ h \nu-\phi.
   \label{eq: A8}
\end{align}

Equations \ref{eq: A4}, \ref{eq: A5}, and \ref{eq: A6} can also be appropriately modified for use in Equation (\ref{eq: A3}).




\section*{References}
\bibliographystyle{unsrt2}
\bibliography{library.bib}

\twocolumngrid
\end{document}